\documentclass[12pt]{article}
\usepackage{amsfonts}
\usepackage{epsfig,amsmath,latexsym,amssymb,color}
\usepackage{amscd}
\usepackage{multirow}
\usepackage{graphicx}
\usepackage{lscape}
\usepackage{bm}

\newcommand{\Remm}[1]{}
\newtheorem{theo}{Theorem}[section]

\newtheorem{rems}[theo]{Remarks}

\newtheorem{model ass}[theo]{Model Assumptions}

\numberwithin{equation}{section}

\begin{document}

\noindent{\Large{\textbf{Dynamic operational risk: modeling
dependence and combining different sources of information}}}

\vspace{1cm}
\noindent\textbf{Gareth W.~Peters  }\\
{\footnotesize {CSIRO Mathematical and Information Sciences, Locked
Bag 17, North Ryde, NSW, 1670, Australia;\\
UNSW Mathematics and Statistics Department, Sydney, 2052,
Australia;\\ email: peterga@maths.unsw.edu.au }}

\vspace{0.5cm}
\noindent\textbf{Pavel V.~Shevchenko (\textit{corresponding author})} \\
{\footnotesize {CSIRO Mathematical and Information Sciences, Sydney,
Locked Bag 17, North Ryde, NSW, 1670, \\
Australia; e-mail: Pavel.Shevchenko@csiro.au}}

\vspace{0.5cm}
\noindent\textbf{Mario V.~W\"{u}thrich}\\
{\footnotesize {ETH Zurich, Department of Mathematics, CH-8092
Zurich, Switzerland; \\email: wueth@math.ethz.ch}}


\begin{center}
\textbf{First Version: 31 October 2007; This Version: 11 April 2009\\
This is a preprint of an article published in \\
The Journal of Operational Risk 4(2), pp. 69-104, 2009
www.journalofoperationalrisk.com}
\end{center}


\begin{abstract}
\noindent In this paper, we model dependence between operational
risks by allowing risk profiles to evolve  stochastically in time
and to be dependent. This allows for a flexible correlation
structure where the dependence between frequencies of different risk
categories and between severities of different risk categories as
well as within risk categories can be modeled. The model is
estimated using Bayesian inference methodology, allowing for
combination of internal data, external data and expert opinion in
the estimation procedure. We use a specialized Markov chain Monte
Carlo simulation methodology known as Slice sampling to obtain
samples from the resulting posterior distribution and estimate the
model parameters.

\vspace{0.5cm} \noindent \textbf{Keywords:} dependence modelling,
copula, compound process, operational risk, Bayesian inference,
Markov chain Monte Carlo, Slice sampling.
\end{abstract}

\pagebreak

\section{Introduction}

Modelling dependence between different risk cells and factors is
an important challenge in operational\ risk (OpRisk) management.
The difficulties of correlation modelling are well known and,
hence, regulators typically take a conservative approach when
considering correlation in risk models. For example, the Basel II
OpRisk regulatory requirements for the
Advanced Measurement Approach, BIS (2006) p.152, states \textquotedblleft \textit{%
Risk measures for different operational risk estimates must be
added for purposes of calculating the regulatory minimum capital
requirement. However, the bank may be permitted to use internally
determined correlations in operational risk losses across
individual operational risk estimates, provided it can demonstrate
to the satisfaction of the national supervisor that its systems
for determining correlations are sound, implemented with
integrity, and take into account the uncertainty surrounding any
such correlation estimates (particularly in
periods of stress). The bank must validate its correlation assumptions using appropriate quantitative and qualitative techniques.}%
\textquotedblright\

The current risk measure specified by regulatory authorities is
Value-at-Risk (VaR) at the 0.999 level for a one year holding
period. In this case simple summation over VaRs corresponds to an
assumption of perfect dependence between risks. This can be very
conservative as it ignores any diversification effects. If the
latter are allowed in the model, capital reduction can be
significant providing a strong incentive to model dependence in the
banking industry. At the same time, limited data does not allow for
reliable estimates of correlations and there are attempts to
estimate these using expert opinions. In such a setting a
transparent dependence model is very important from the perspective
of model interpretation, understanding of model sensitivity and with
the aim of minimizing possible model risk. However, we would also
like to mention that VaR is not a coherent risk measure, see
Artzner, Delbaen, Eber and Heath (1999). This means that in
principal dependence modelling could also increase VaR, see
Embrechts, Ne\v{s}lehov\'{a} and W\"{u}thrich (2009) and Embrechts,
Lambrigger and W\"{u}thrich (2009).

Under Basel II requirements, the financial institution intending to
use the Advanced Measurement Approach (AMA) for quantification of
OpRisk should demonstrate accuracy of the internal model within 56
risk cells (eight business lines times seven event types). To meet
regulatory requirements, the model should make use of internal data,
relevant external data, scenario analysis and factors reflecting the
business environment and internal control systems. The definition of
OpRisk, Basel II requirements and the possible Loss Distribution
Approach for AMA were discussed widely in the literature, see e.g.
Cruz (2004), Chavez-Demoulin, Embrechts and Ne\v{s}lehov\'{a}
(2006), Frachot, Moudoulaud and Roncalli (2004), Shevchenko (2009).
It is more or less widely accepted that under the Loss Distribution
Approach of AMA Basel II requirements, the banks should quantify
distributions for frequency and severity of OpRisk for each business
line and event type over a one year time horizon. These are combined
into an annual loss distribution for the bank top level (as well as
business lines and event types if required) and the bank capital
(unexpected loss) is estimated using the 0.999 quantile of the
annual loss distribution. If the severity and frequency distribution
parameters are known, then the capital estimation can be
accomplished using different techniques. In the case of single risks
there are: hybrid Monte Carlo approaches, see Peters, Johansen and
Doucet (2007); Panjer Recursions, see Panjer (1981); integration of
the characteristic functions, see Luo and Shevchenko (2009); Fast
Fourier Transform techniques, see e.g. Embrechts and Frei (2009),
Temnov and Warnung (2008). To account for parameter uncertainty, see
Shevchenko (2008), and in multivariate settings Monte Carlo methods
are typically used.

The commonly used model for an annual loss in a risk cell (business
line/event type) is a compound random variable,%
\begin{equation}
Z_{t}^{\left( j\right) }=\sum\limits_{s=1}^{N_{t}^{\left( j\right)
}}X_{s}^{\left( j\right) }\left( t\right).  \label{Model1}
\end{equation}%
Here $t=1,2,\ldots,T,T+1$ in our framework is discrete time (in
annual units) with $T+1$ corresponding to the next year. The upper
script $j$ is used to identify the risk cell. The annual number of
events $N_{t}^{( j) }$ is a random variable distributed according to a frequency counting distribution $%
P^{(j) }( \cdot |{\lambda }_{t}^{(j)}) $, typically Poisson, which
also depends on time dependent parameter(s) ${\lambda }_{t}^{(j)}$.
The severities in year $t$ are represented by
random variables $X_{s}^{(j)}(t)$, $s\ge1$, distributed according to a severity distribution $F^{(j)}(\cdot |{\psi }%
_{t}^{(j)})$, typically lognormal, Weibull or generalized Pareto
distributions with parameter(s) ${\psi }_{t}^{(j)}$. Note, the
index $j$ on the distributions $P^{(j)}$ and $F^{(j)}$ reflects
that distribution type can be different for different risks, for
simplicity of notation we shall
omit this $j$, using $P(\cdot |{\lambda }_{t}^{(j)})$ and $F(\cdot |{\psi}%
_{t}^{(j)})$, hereafter. The variables ${\lambda }_{t}^{(j)}$ and ${\psi}%
_{t}^{(j)}$ generically represent distribution (model) parameters of the $%
j^{th}$ risk that we refer hereafter to as the risk profiles. Typically, it
is assumed that given ${\lambda}_{t}^{(j)}$ and ${\psi }_{t}^{(j)}$, the
frequency and severities of the $j^{th} $ risk are independent and the
severities within the $j^{th}$ risk are independent too. The total bank's
loss in year $t$ is calculated as
\begin{equation}
Z_{t}=\sum\limits_{j=1}^{J}Z_{t}^{\left( j\right) },
\end{equation}%
where formally for OpRisk under the Basel II requirements $J=56$
(seven event types times eight business lines). However, this may
differ depending on the financial institution and type of problem.

Conceptually under model (\ref{Model1}), the dependence between
the annual losses $Z_{t}^{\left( j\right) }$ and $Z_{t}^{\left(
i\right) }, i \neq j,$ can be introduced in several ways. For
example via:

\begin{itemize}
\item Modelling dependence between frequencies $N_{t}^{\left( j\right) }$
and $N_{t}^{\left( i\right) }$ directly through e.g. copula
methods, see e.g. Frachot, Roncalli and Salomon (2004), Bee (2005)
and Aue and Klakbrener (2006) or common shocks, see e.g. Lindskog
and McNeil (2003), Powojowski, Reynolds and Tuenter (2002). We
note that the use of copula methods, in the case of discrete
random variables, needs to be done with care. The approach of
common shocks is proposed as a method to model events affecting
many cells at the same time. Formally, this leads to dependence
between frequencies of the risks if superimposed with cell
internal events. One can introduce the dependence between event
times of different risks, e.g. the $1^{st}$ event time of the $j^{th}$ risk correlated to the $%
1^{st}$ event time of the $i^{th}$ risk, etc., but it can be
problematic to interpret such a model.

\item Considering dependence between severities (e.g. the first loss amount
of the $j^{th}$ risk is correlated to the first loss of the
$i^{th}$ risk, second loss in the $j^{th}$ risk is correlated to
second loss in the $i^{th}$ risk, etc), see e.g. Chavez-Demoulin,
Embrechts and Ne\v{s}lehov\'{a} (2006). This can be difficult to
interpret especially when one considers high frequency versus low
frequency risks.

\item Modelling dependence between annual losses directly via copula methods,
see Giacometti, Rachev, Chernobai and Bertocchi (2008), B\"{o}cker
and Kl\"{u}ppelberg (2008) and Embrechts and Puccetti (2008).
However, this may create irreconcilable problems with modelling
insurance for OpRisk that directly involves event times.
Additionally, it will be problematic to quantify these
correlations using historical data, and the LDA model
(\ref{Model1}) will loose its structure. Though one can consider
dependence between losses aggregated over shorter periods.
\end{itemize}

\noindent In this paper, we assume that all risk profiles are stochastically
evolving in time. That is we model risk profiles $\bm{\lambda }_{t}=\left( {%
\lambda }_{t}^{\left( 1\right) },\ldots,{\lambda }_{t}^{\left(
J\right) }\right) $ and $\bm{\psi }_{t}=\left( {\psi }_{t}^{\left(
1\right) },\ldots,{\psi }_{t}^{\left( J\right)
}\right) $ by random variables $\bm{\Lambda }_{t}=\left( {\Lambda }%
_{t}^{\left( 1\right) },\ldots,{\Lambda }_{t}^{\left( J\right) }\right) $ and $%
\bm{\Psi }_{t}=\left( {\Psi }_{t}^{\left( 1\right) },\ldots,{\Psi
}_{t}^{\left( J\right) }\right) ,$ respectively. We introduce
dependence between risks by
allowing dependence between their risk profiles $\bm{\Lambda }_{t}$ and $%
\bm{\Psi }_{t}$. Note that, independence between frequencies and
severities in (\ref{Model1}) is conditional on risk profiles $\left(
\bm{\Lambda }_{t},\bm{\Psi }_{t}\right)$ only. Additionally we
assume, for the sake of simplicity, that all risks are independent
conditional on risk profiles.

Stochastic modelling of risk profiles may appeal to intuition. For
example consider the annual number of events for the $j^{th} $
risk modelled as random variables from Poisson distribution $%
Poi\left( {\Lambda }_{t}^{\left( j\right) }={\lambda }_{t}^{\left(
j\right) }\right) $. Conditional on ${\Lambda }_{t}^{\left(
j\right) }$, the expected number of events per year is ${\Lambda
}_{t}^{\left( j\right) }$. The latter is not only different for
different banks and different risks but also changes from year to
year for a risk in the same bank. In general, the evolution of
${\Lambda }_{t}^{\left( j\right) }$, can be modelled as having
deterministic (trend, seasonality) and stochastic components. In
actuarial mathematics this is called a mixed Poisson model. For
simplicity, in this paper, we assume that ${\Lambda }_{t}^{\left(
j\right) }$ is purely stochastic and distributed according to a
Gamma distribution.

Now consider a sequence $\left(
\bm{\Lambda}_{1},\bm{\Psi}_{1}\right), \ldots, \left( \bm{\Lambda
}_{T+1},\bm{\Psi }_{T+1}\right)$. It is naive to assume that risk
profiles of all risks are independent. Intuitively these are
dependent, for example, due to changes in politics, regulations,
law, economy, technology (sometimes called drivers or external
risk factors) that jointly impact on many risk cells at each time
instant. In this paper we focus on dependence between risk
profiles.

We begin by presenting the general model and then we perform
analysis of relevant properties of this model in a bivariate risk
setting. Next, we demonstrate how to perform inference under our
model by adopting a Bayesian approach that allows one to combine
internal data with expert opinions and external data. We consider
both the single risk and multiple risk settings for the example of
modelling claims frequencies. Then we present an advanced
simulation procedure utilizing a class of Markov chain Monte Carlo
(MCMC) algorithms which allow us to sample from the posterior
distributions developed. Finally, we demonstrate the performance
of both the model and the simulation procedure in several
examples, before finishing with a discussion and conclusions.

The main objective of the paper is to present the framework we
develop for the multivariate problem and to demonstrate estimation
in this setting. Application of real data is the subject of further
research. To clarify notation, we shall use upper case symbols to
represent random variables, lower case symbols for their
realizations and bold for vectors.

\section{Model}

\begin{model ass}
Consider $J$ risks each with a general model (\ref{Model1}) for the annual
loss in year $t$, $Z_{t}^{\left( j\right) }$, and each modelled by severity $%
X_{s}^{\left( j\right) }(t)$ and frequency $N_{t}^{\left( j\right) }$. The
frequency and severity risk profiles are modelled by random vectors $%
\bm{\Lambda }_{t}=({\Lambda }_{t}^{(1)},\ldots,{\Lambda
}_{t}^{\left( J\right)})$ and $\bm{\Psi }_{t}=({\Psi }_{t}^{\left( 1\right) },\ldots,{\Psi }%
_{t}^{\left( J\right) })$ respectively and parameterized by risk
characteristics $\bm{\theta}_{\Lambda }=$ $(\theta _{\Lambda
}^{\left(1\right) },\ldots,\theta_{\Lambda }^{\left( J\right) }$) and $\bm{\theta}%
_{\Psi }=(\theta _{{\Psi }}^{\left( 1\right) },\ldots,\theta
_{{\Psi }}^{\left( J\right) })$ correspondingly. Additionally, the
dependence between risk profiles is parameterized by
$\bm{\theta}_{\rho}$. Assume that, given $\bm{\theta}=(\bm{\theta}
_{\Lambda },\bm{\theta}_{\Psi },\bm{\theta}_{\rho})$:

\begin{enumerate}
\item The random vectors,
\begin{eqnarray*}
&&\left( \bm{\Psi }_{1},\bm{\Lambda }_{1},N_{1}^{\left( j\right)
},X_{s}^{(j)}\left( {1}\right) ;j=1,\ldots ,J, s\ge1 \right) \\
&&\vdots \\
&&\left( \bm{\Psi }_{T+1},\bm{\Lambda }_{T+1},N_{T+1}^{\left(
j\right) },X_{s}^{(j)}\left( {T+1}\right) ;j=1,\ldots ,J, s\ge1
\right)
\end{eqnarray*}%
are independent. That is, given $\bm{\theta}$, between different
years, the risk profiles for frequencies and severities as well as
the number of losses and actual losses are independent.

\item The vectors $\left( \bm{\Psi}_{1},\bm{\Lambda}_{1}\right),\ldots,\left( \bm{\Psi}_{T+1},\bm{\Lambda}_{T+1}\right) $ are
\textit{i.i.d.} from a joint distribution with marginal
distributions ${\Lambda }_{t}^{(j)}\sim G\left( \cdot |{\theta
}_{{\Lambda }}^{(j)}\right) $, ${\Psi }_{t}^{(j)}\sim
H\left( \cdot |{\theta }_{{\Psi }}^{\left( j\right) }\right) $ and $2J$%
-dimensional copula $C(\cdot|\bm{\theta}_{\rho})$.

\item Given $\bm{\Lambda }_{t}=\bm{\lambda }_{t}$ and $\bm{\Psi }_{t}=%
\bm{\psi}_{t}$: the compound random variables $Z_{t}^{\left(
1\right) },\ldots ,Z_{t}^{\left( J\right) }$ are independent with
$N_{t}^{\left( j\right) }$ and $X_{1}^{\left( j\right) }\left(
t\right), X_{2}^{\left( j\right) }\left( t\right),\ldots$ independent; frequencies $%
N_{t}^{\left( j\right) }\sim P\left( \cdot |{\lambda }_{t}^{\left( j\right)
}\right) $; and severities $X_{s}^{(j)}\left( t\right) \overset{i.i.d.}{\sim }%
F\left( \cdot |{\psi }_{t}^{\left( j\right) }\right) ,s\ge1.$
\end{enumerate}

\label{GeneralModelAssumptions}
\end{model ass}

\noindent Calibration of the above model requires estimation of
$\bm{\theta}$. A thorough discussion about the interpretation and role of $%
\bm{\theta}$ is provided in Section 4, where it will be treated
within a Bayesian framework as a random variable $\bm{\Theta}$ to
incorporate expert opinions and external data into the estimation
procedure. Also note that for simplicity of notation, we assumed
one severity risk profile $\Psi _{t}^{(j)} $ and one frequency
risk profile $\Lambda _{t}^{(j)}$ per risk - extension is trivial
if more risk profiles are required to model risk.

~

\noindent \textbf{Copula models.} To define the above model, a
copula function $C(\cdot )$ should be specified to model dependence
between the risk profiles. For a description of copulas in the
context of financial risk modelling see McNeil, Frey and Embrechts
(2005). In general, a copula is a $d$-dimensional
multivariate distribution on $[0,1]^{d}$ with uniform marginal distributions. Given a copula function $%
C(u_{1},\ldots ,u_{d})$, the joint distribution of rvs
$Y_{1},\ldots ,Y_{d}$ with marginal distributions
$F_{1}(y_{1}),\ldots ,F_{d}(y_{d})$ can be constructed as
\begin{equation}
F(y_{1},\ldots ,y_{d})=C(F_{1}(y_{1}),\ldots ,F_{d}(y_{d})).
\label{CopulaDefinition}
\end{equation}%
\noindent A well known theorem due to Sklar, published in 1959, says that
one can always find a unique copula $C(\cdot )$ for a joint distribution
with given continuous marginals. Note that in the case of discrete
distributions this copula may not be unique. Given (\ref{CopulaDefinition}),
the joint density can be written as
\begin{equation}
f(y_{1},\ldots ,y_{d})=c(F_{1}(y_{1}),\ldots
,F_{d}(y_{d}))\prod\limits_{i=1}^{d}f_{i}(y_{i}).
\label{CopulaDensityDefinition}
\end{equation}

\noindent where $c(\cdot)$ is a copula density and
$f_1(y_1),\ldots,f_d(y_d)$ are marginal densities. In this paper,
for illustration purposes we consider the Gaussian, Clayton and
Gumbel copulas (Clayton and Gumbel copulas belong to the so-called
family of the Archimedean copulas):

\begin{itemize}
\item \textbf{Gaussian copula:}
\begin{equation}
c\left( u_{1},\ldots ,u_{d}|\Sigma\right)
=\frac{f_N^{\Sigma}\left( F_{N}^{-1}(u_{1}) ,\ldots
,F_{N}^{-1}(u_{d}) \right)}{\prod\limits_{i=1}^{d} f_N \left(
F_{N}^{-1}(u_{i}) \right) },  \label{GaussianCopula}
\end{equation}
where $F_N(\cdot)$ and $f_N(\cdot)$ are the standard Normal distribution and
its density respectively and $f_N^{\Sigma}(\cdot)$ is a multivariate Normal
density with zero means, unit variances and correlation matrix $\Sigma$.

\item \textbf{Clayton copula}:
\begin{equation}
c\left( u_{1},\ldots ,u_{d}|\rho\right) =\left(
1-d+\sum\limits_{i=1}^{d}\left( u_{i}\right) ^{-\rho }\right)
^{-d-\frac{1}{\rho }}\prod\limits_{i=1}^{d} \left( \left(
u_{i}\right) ^{-\rho -1}\left\{ \left( i-1\right) \rho +1\right\}
\right),  \label{ClaytonCopula}
\end{equation}
where $\rho > 0$ is a dependence parameter.
\item \textbf{Gumbel copula}:
\begin{eqnarray}
c\left( u_{1},\ldots ,u_{d}|\rho\right) &=&\frac{\partial
^{d}}{\partial u_{1}\ldots \partial u_{d}}C\left( u_{1},\ldots
,u_{d}|\rho\right) ,\text{ } \\
C\left( u_{1},\ldots ,u_{d}|\rho\right) &=&\exp \left\{ -\left(
\sum\nolimits_{i=1}^{d}\left( -\log \left( u_{i}\right) \right)
^{\rho }\right) ^{\frac{1}{\rho }}\right\}, \label{GambelCopula}
\end{eqnarray}
where $\rho \geq 1$ is a dependence parameter.

In the bivariate case the explicit expression for Gumbel copula is given by
\begin{eqnarray*}
c\left( u_{1},u_{2}|\rho\right) &=&\frac{\partial ^{2}}{\partial
u_{1}\partial
u_{2}}C\left( u_{1},u_{2}|\rho\right) \\
&=&C\left( u_{1},u_{2}|\rho\right) u_{1}^{-1}u_{2}^{-1}\left[ \sum%
\nolimits_{i=1}^{2}\left( -\log \left( u_{i}\right) \right) ^{\rho }\right]
^{2\left( \frac{1}{\rho }-1\right) }\left[ \log \left( u_{1}\right) \log
\left( u_{2}\right) \right] ^{\rho -1} \\
&&\times \left[ 1+\left( \rho -1\right) \left[ \sum\nolimits_{i=1}^{2}\left(
-\log \left( u_{i}\right) \right) ^{\rho }\right] ^{-\frac{1}{\rho }}\right]
.
\end{eqnarray*}
\end{itemize}

An important difference between these three copulas is that they
each display different tail dependence properties. The Gaussian
copula has no upper or lower tail dependence, the Clayton copula
produces lower tail dependence, whereas the Gumbel copula produces
upper tail dependence, see McNeil, Frey and Embrechts (2005).

~

\noindent \textbf{Common factor models.} The use of common
(systematic) factors is useful to identify dependent risks and to
reduce the number of required correlation coefficients that must be
estimated. For example, assuming a Gaussian copula between risk
profiles, consider one common factor $\Omega _{t}$ affecting all
risk profiles as follows
\begin{eqnarray}
Y_{t}^{(i)} &=&\rho _{i}\Omega _{t}+\sqrt{1-\rho _{i}^{2}}%
W_{t}^{(i)},i=1,\ldots ,2J;  \notag \\
\Lambda _{t}^{(j)} &=&G^{-1}(F_{N}(Y_{t}^{(j)})|\theta _{\Lambda
}^{(j)}),\Psi _{t}^{(j)}=H^{-1}(F_{N}(Y_{t}^{(j+J)})|\theta _{\Psi
}^{(j)}),j=1,\ldots ,J,
\end{eqnarray}%
where $W_{t}^{(1)},\ldots,W_{t}^{(2J)}$ and $\Omega _{t}$ are iid
from the standard Normal distribution and all rvs are independent
between different time steps $t$. Given $\Omega _{t}$, all risk
profiles are independent but unconditionally the risk profiles are
dependent if the corresponding $\rho _{i}$ are nonzero. In this
example, one should identify $2J$ correlation parameters $\rho_i$
only instead of $J(J-1)/2$ parameters of the full correlation
matrix. Often, common factors are unobservable and practitioners use
generic intuitive definitions such as: changes in political, legal
and regulatory environments, economy, technology, system security,
system automation, etc. Several external and internal factors are
typically considered. The factors may affect the frequency risk
profiles (e.g. system automation), the severity risk profiles (e.g.
changes in legal environment) or both the frequency and severity
risk profiles (e.g. system security). For more details on the use
and identification of the factor models, see Section 3.4 in McNeil,
Frey and Embrechts (2005); also, see Sections 5.3 and 7.4 in
Marshall (2001) for the use in the operational risk context.

In general, a copula can be introduced between all risk profiles.
Though, for simplicity, in the simulation examples below,
presented for two risks, we consider dependence between severities
and frequencies separately. Also, in this paper, the estimation
procedure is presented for frequencies only. The actual procedure
can be extended in the same manner as presented to severities but
it is the subject of further work.

\section{{\protect\LARGE \textbf{Simulation Study - Bivariate Case}}}

We start with a bivariate model, where we study the strength of dependence
at the annual loss level obtained through dependence in risk profiles, as
discussed above. We consider two scenarios. The first involves independent
severity risk profiles and dependent frequency risk profiles. The second
involves dependence between the severity risk profiles and independence
between the frequency profiles. In both scenarios, we consider three
bivariate copulas (Gaussian, Clayton and Gumbel copulas (\ref{GaussianCopula}%
)-(\ref{GambelCopula})) denoted as $C(u_{1},u_{2}|\rho)$ and
parameterized by one parameter $\rho$ which controls the degree of
dependence. In the case of Gaussian copula, $\rho$ is a
non-diagonal element of correlation matrix $\Sigma $ in
(\ref{GaussianCopula}).

~

\noindent \textbf{Bivariate model for risk profiles.} We assume that
Model Assumptions 2.1 are fullfilled for the aggregated losses
\begin{equation*}
Z_{t}^{\left( 1\right) }=\sum\limits_{s=1}^{N_{t}^{\left( 1\right)
}}X_{s}^{(1)}\left( t\right) \text{ \ \ and \ }Z_{t}^{\left(
2\right) }=\sum\limits_{s=1}^{N_{t}^{\left( 2\right)
}}X_{s}^{(2)}\left( t\right) \text{.}
\end{equation*}%
As marginals, for $j=1,2$ we choose:

\begin{itemize}
\item $N_{t}^{(j)}\sim Poi\left( \lambda _{t}^{(j)}\right)$ and $X_{s}^{(j)}(t)\overset{i.i.d.}{\sim }%
LN\left( \psi^{(j)}_t,{\sigma}^{(j)}\right) $, $%
s\ge1$.

\item $\Lambda _{t}^{(j)}\sim \Gamma \left( \alpha _{\Lambda }^{\left(
j\right) },\beta _{\Lambda }^{\left( j\right) }\right) $, $\Psi
^{(j)}_t\sim N\left( \mu_{\Psi}^{(j)},\omega_{\Psi}^{(j)}\right) $.
\end{itemize}

\noindent Here, $\Gamma \left( \alpha ,\beta \right) $ is a Gamma
distribution with mean $\alpha /\beta $ and variance $\alpha/\beta^2$, $%
N\left( \mu ,\sigma \right) $ is a Gaussian distribution with mean $\mu $
and standard deviation $\sigma $, and $LN^{{}}\left( \mu ,\sigma \right) $
is a lognormal distribution.

In analyzing the induced dependence between annual losses, we consider two
scenarios:

\begin{itemize}
\item Scenario 1: $\Lambda _{t}^{\left( 1\right) }$ and $\Lambda
_{t}^{\left( 2\right) }$ are dependent via copula $C(u_1,u_2|\rho)$
while $\Psi_{t}^{(1)}$ and $\Psi_{t}^{(2)}$ are independent.

\item Scenario 2: $\Psi_{t}^{(1)}$ and $\Psi_{t}^{(2)}$ are
dependent via copula $C(u_1,u_2|\rho)$ while $\Lambda_{t}^{\left(
1\right) }$ and $\Lambda_{t}^{\left( 2\right) }$ are independent.
\end{itemize}

\noindent Here, parameter $\rho$ corresponds to $\theta_{\rho}$ in
Model Assumptions 2.1. The simulation of the annual losses when
risk profiles are dependent via a copula can be accomplished as
shown in Appendix A. Utilizing this procedure, we examine the
strength of dependence between the annual losses if there is a
dependence between the risk profiles. In the next sections we will
demonstrate the Bayesian inference model and associated
methodology to perform estimation of the model parameters. Here,
we assume the parameters are known \textit{a priori} with the
following values used in our specific example:

\begin{itemize}
\item $\alpha_{\Lambda }^{\left( j\right) }=5,\beta _{\Lambda }^{\left(
j\right) }=0.1,\mu_{\Psi}^{\left( j\right) }=2,\omega_{\Psi
}^{\left( j\right) }=0.4,\sigma^{(j)}=1; j=1,2$
\end{itemize}

\begin{figure}[ptb]
\centerline{%
\includegraphics[width=5.0in,height=4.0in]{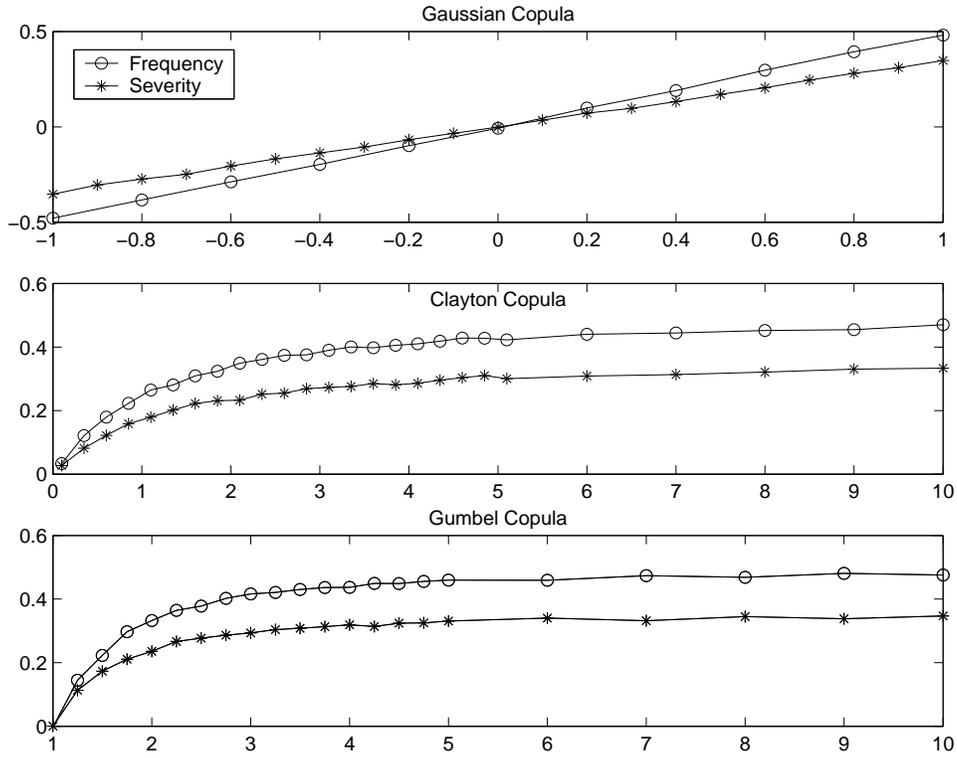}}
\caption{{\protect\footnotesize {Spearman's rank correlation, $\protect\rho %
_{SR}\left( Z^{(1)},Z^{(2)}\right) $, between the annual losses vs
copula
parameter $\protect\rho $, also see Section 3. \ ($\circ $) $\protect\rho %
_{SR}\left( Z^{(1)},Z^{(2)}\right) $ vs copula parameter
$\protect\rho $ between
frequency risk profiles $\Lambda^{(1)}_{t}$ and $\Lambda^{(2)}_{t}$; ($\ast $%
) $\protect\rho _{SR}\left( Z^{(1)},Z^{(2)}\right) $ vs copula parameter $%
\protect\rho $ between severity risk profiles $\Psi_{t}^{(1)}$ and
$\Psi_{t}^{(2)}$.}}} \label{fig1}
\end{figure}

\noindent These parameters correspond to $\bm{\theta}_{\Lambda}$ and
$\bm{\theta}_{\Psi}$ in Model Assumptions 2.1. In Figure \ref{fig1},
we present three cases where $C\left( \cdot|\rho\right) $ is a
Gaussian, Clayton or Gumbel copula under both scenario 1 and
scenario 2. In each of these examples we vary the parameter of the
copula model $\rho$ from weak to strong dependence. The annual
losses are not Gaussian distributed and to measure the dependence
between the annual losses we use a non-linear rank correlation
measure, Spearman's rank correlation, denoted by $\rho _{SR}(
Z_{t}^{(1) },Z_{t}^{(2)}) $. The Spearman's rank correlation between
the annual losses was estimated using $10,000$ simulated years for
each value of $\rho$. In these and other numerical experiments we
conducted, the range of possible dependence between the annual
losses of different risks induced by the dependence between risk
profiles is very wide and should be flexible enough to model
dependence in practice. Note, the degree of induced correlation can
be further extended by working with more flexible copula models at
the expense of estimation of a larger number of model parameters.

\section{Bayesian Inference: combining different data sources}

In this section we estimate the model introduced in Section 2 using
a Bayesian inference method. To achieve this we must consider that
the requirements of Basel II AMA (see BIS, p.152) clearly state
that: \textit{"Any operational risk measurement system must have
certain key features to meet the supervisory soundness standard set
out in this section. These elements must include the use of internal
data, relevant external data, scenario analysis and factors
reflecting the business environment and internal control systems".}
Hence, Basel II requires that OpRisk models include use of several
different sources of information. We will demonstrate that to
satisfy such requirements it is important that methodology such as
the one we develop in this paper be considered in practice to ensure
one can soundly combine these different data sources.

It is widely recognized that estimation of OpRisk frequency and
severity distributions cannot be done solely using historical
data. The reason is the limited ability to predict future losses
in a banking environment which is constantly changing. Assume that
a new policy was introduced in a financial institution with the
intention of reducing an OpRisk loss. This cannot be captured in a
model based solely on historical loss data.

For the above reasons, it is very important to include Scenario
Analysis (SA) in OpRisk modelling. SA is a process undertaken by
banks to identify risks; analyze past events experienced
internally and jointly with other financial institutions including
near miss losses; consider current and planned controls in the
banks, etc. Usually, it involves surveying of experts through
workshops. A template questionnaire is developed to identify
weaknesses, strengths and other factors. As a result an imprecise,
value driven quantitative assessment of risk frequency and
severity distributions is obtained. On its own, SA is very
subjective and we argue it should be combined (supported) by
actual loss data analysis. It is not unusual that correlations
between risks are attempted to be specified by experts in the
financial institution, typically via SA.

External loss data is also an important source of information which should
be incorporated into modelling. There are several sources available to
obtain external loss data, for a discussion on some of the data related
issues associated with external data see Peters and Teruads (2007).

Additionally, the combination of expert opinions with internal and
external data is a difficult problem and complicated ad-hoc
procedures are used in practice. Some prominent risk professionals
in industry have argued that statistically consistent combining of
these different data sources is one of the most pertinent and
challenging aspects of OpRisk modelling. It was quoted in Davis
(2006) \textit{"Another big challenge for us is how to mix the
internal data with external data; this is something that is still a
big problem because I don't think anybody has a solution for that at
the moment"} and \textit{"What can we do when we don't have enough
data [$\cdots$] How do I use a small amount of data when I can have
external data with scenario generation? [$\cdots$] I think it is one
of the big challenges for operational risk managers at the
moment."}. Using the methodology that we develop in this paper, one
may combine these data sources in a statistically sound approach,
addressing these important practical questions that practitioners
are facing under Basel II AMA.

Bayesian inference methodology is well suited to combine different
data sources in OpRisk, for example see Shevchenko and
W\"{u}thrich (2006). A closely related credibility theory toy
example was considered in B\"{u}hlmann, Shevchenko and
W\"{u}thrich (2007). We also note that in general questions of
Bayesian model choice must be addressed, adding to this there is
the additional complexity that estimation of the required
posterior distributions will typically require MCMC, see Peters
and Sisson (2006).

A Bayesian model to combine three data sources (internal data,
external data and expert opinion) for the case of a single risk
cell was presented in Lambrigger, Shevchenko and W\"{u}thrich
(2007). In this paper we extend this approach to the case of many
risk cells with the dependence between risks introduced as in
Section 2. Hereafter, for illustrative purposes we restrict to
modelling frequencies only.

Hence, our objective will be to utilise Bayesian inference to
estimate the parameters of the model through the combination of
expert opinions and observed loss data (internal and external).

We note that as part of this Bayesian model formualtion an
information flow can be incorporated into the model. This could be
introduced in many forms. The most obvious example involves
incorporation of new data from actual observed losses. However, we
stress that more general ideas are possible. For example, if new
information becomes available (new policy introduced, etc) then
experts can update their prior distributions to incorporate this
information into the model.

Additionally, under a Bayesian model we note that SA could naturally
form part of a subjective Bayesian prior elicitation procedure, see
O'Hagan (2006).

\subsection{Modelling frequencies for a single risk cell}

Here we follow the Lambrigger, Shevchenko and W\"{u}thrich (2007)
approach to combine different data sources for one risk cell in
the case of the Model Assumptions \ref{GeneralModelAssumptions}.

Define a model in which every risk cell of a financial company
$j\in \{1,\ldots ,J\}$ is characterized by a risk characteristic
$\Theta_{\Lambda}^{(j)}$
that describes the frequency risk profile $\Lambda _{t}^{(j)}$ in risk cell $%
j$. This $\Theta_{\Lambda}^{(j)}$ represents a vector of unknown
distribution parameters of risk profile $\Lambda^{(j)}_{t}$. The
true value of $\Theta_{\Lambda}^{(j)}$ is not known and modelled
as a random variable. \textit{A priori}, before having any company
specific information, the prior distribution of $\Theta_{\Lambda}
^{(j)}$ is based on external data only. Our aim then is to specify
the distribution of $\Theta_{\Lambda}^{(j)}$ when we have company
specific information about risk cell $j$ such as observed losses
and expert opinions. This is achieved by developing a Bayesian
model and numerical estimation of relevant quantities is performed
via MCMC methods. For simplicity, in this section, we drop the
risk cell specific superscript $j$ since we concentrate on
modelling frequencies for single risk cell $j$, where
$\Theta_{\Lambda}^{(j)}$ is a scalar $\Theta_{\Lambda} $ and all
other parameters are assumed known.

\begin{model ass}
\textrm{Assume that risk cell }$j$\textrm{\ has a fixed, deterministic
volume $V$ (i.e. number of transactions, etc.). }

\begin{enumerate}
\item \textrm{The risk characteristics $\Theta_{\Lambda} $ of risk cell }$j$\textrm{\
has prior distribution: $\Theta_{\Lambda} \sim \Gamma (a,b)$ for
given parameters $a>0$ and $b>0$.}

\item \textrm{Given $\Theta_{\Lambda} =\theta_{\Lambda} $, $(\Lambda _{1},N_{1}),\ldots
,(\Lambda _{T+1},N_{T+1})$ are i.i.d.~and the intensity of events of year $%
t\in \{1,\ldots ,T+1\}$ has conditional marginal distribution
$\Lambda
_{t}\sim \Gamma (\alpha ,\alpha /\theta_{\Lambda} )$ for a given parameter $\alpha >0$%
. }

\item \textrm{Given $\Theta =\theta_{\Lambda} $ and $\Lambda _{t}=\lambda _{t}$, the frequencies $N_{t}\sim Poi(V\lambda _{t})$. }

\item \textrm{The financial company has $K$ expert opinions $%
\Delta _{k}$, $k= 1,\ldots ,K$ about }$\Theta_{\Lambda} $. Given $%
\Theta_{\Lambda} =\theta_{\Lambda} $, $\Delta _{k}$ and $(\Lambda
_{t},N_{t})$ are independent for all $k$ and $t$\textrm{, and
$\Delta _{1},\ldots,\Delta_{K}$ are i.i.d.~with $\Delta _{k}\sim
\Gamma (\xi ,\xi /\theta_{\Lambda} )$. }
\end{enumerate}
\end{model ass}

\begin{rems}
~
\begin{itemize}
\item \textnormal{In items 1) and 2) we choose a gamma distribution for
the underlying parameters. Often, the available data is not
sufficient to support such a choice. In such cases, in actuarial
practice, one often chooses a gamma distribution. A gamma
distribution is neither conservative nor aggressive and it has the
advantage that it allows for transparent model interpretations. If
other distributions are more appropriate then, of course, one should
replace the gamma assumption. This can easily be done in our
simulation methodology.}

\item \textnormal{Given that $\Theta_{\Lambda} \sim \Gamma (a,b)$, $E\left[ \Theta_{\Lambda} %
\right] =a/b$ and $Var\left( \Theta_{\Lambda} \right) =a/b^{2}$.
These are the prior two moments of the underlying risk
characteristics }\textrm{$\Theta_{\Lambda} .$} \textnormal{The prior
can be determined by external data (or regulator). In general,
parameters $a$ and $b$ can be estimated by the maximum likelihood
method using the data from all banks.}

\item \textnormal{Note that we have for the first moments
\begin{eqnarray}
E\left[ \left. \Lambda _{t}\right\vert \Theta_{\Lambda} \right] &=&\Theta_{\Lambda} ,\text{ }E%
\left[ \Lambda _{t}\right] =\frac{a}{b},\text{ }E\left[ \left.
N_{t}\right\vert \Theta_{\Lambda} ,\Lambda _{t}\right] =V~\Lambda _{t},  \notag \\
E\left[ \left. N_{t}\right\vert \Theta_{\Lambda} \right]
&=&V~\Theta_{\Lambda} ,\text{ }E\left[ N_{t}\right]
=V~\frac{a}{b}.  \notag
\end{eqnarray}%
The second moments are given by
\begin{eqnarray}
Var\left( \left. \Lambda _{t}\right\vert \Theta_{\Lambda} \right)
&=&\alpha
^{-1}~\Theta_{\Lambda} ^{2},~~\text{ }Var\left( \Lambda _{t}\right) =\alpha ^{-1}~%
\frac{a^{2}}{b^{2}}+(\alpha ^{-1}+1)~\frac{a}{b^{2}},  \notag \\
Var\left( \left. N_{t}\right\vert \Theta_{\Lambda} ,\Lambda
_{t}\right) &=&V~\Lambda _{t},\text{ }~~Var\left( \left.
N_{t}\right\vert \Theta_{\Lambda} \right) =V~\Theta_{\Lambda}
+V^{2}~\alpha ^{-1}~\Theta_{\Lambda} ^{2}, \\
Var\left( N_{t}\right) &=&V~\frac{a}{b}+V^{2}~\alpha ^{-1}~\frac{a^{2}}{b^{2}%
}+V^{2}~(\alpha ^{-1}+1)~\frac{a}{b^{2}}.  \notag
\end{eqnarray}%
For model interpretation purposes, consider the results for the
coefficient of variation (CV), a convenient dimensionless measure of
uncertainty commonly used in the insurance industry:
\begin{equation}
\lim_{V\rightarrow \infty }CV^{2}\left( \left. N_{t}\right\vert
\Theta_{\Lambda} \right) =\lim_{V\rightarrow \infty
}\frac{Var\left( \left. N_{t}\right\vert \Theta_{\Lambda} \right)
}{E^{2}\left[ \left. N_{t}\right\vert \Theta_{\Lambda} \right]
}=\alpha ^{-1}>0,  \label{div1}
\end{equation}%
and
\begin{equation}
\lim_{V\rightarrow \infty }CV^{2}\left( N_{t}\right)
=\lim_{V\rightarrow \infty }\frac{Var\left( N_{t}\right)
}{E^{2}\left[ N_{t}\right] }=\alpha ^{-1}+\left( \alpha
^{-1}+1\right) ~a^{-1}>0.  \label{div2}
\end{equation}%
That is, our model makes perfect sense from a practical perspective.
Namely, as volume increases, $V\rightarrow \infty $, there always
remains a non-diversifiable element, see \eqref{div1} and
\eqref{div2}. This is exactly what has been observed in practice and
what regulators require from internal models. Note, if we model
$\Lambda_t$ as constant and known then $\lim_{V\rightarrow \infty
}CV^{2}\left( N_{t}|\Lambda_t\right) \rightarrow 0$.}

\item \textnormal{Contrary to the developments in Lambrigger, Shevchenko and
W\"{u}thrich (2007), where the intensity $\Lambda_t$ was constant
overtime, now $\Lambda_t$ is a stochastic process. From a
practical point of view, it is not plausible that the intensity of
the annual counts is constant over time. In such a setting
parameter risks completely vanish if we have infinitely many
observed years or infinitely many expert opinions, respectively
(see Theorem 3.6 (a) and (c) in Lambrigger, Shevchenko and
W\"{u}thrich (2007)). This is because $\ \Lambda _{t}$ can then be
perfectly forecasted. In the present model, parameter risks will
also decrease with increasing information. As we gain information
the posterior standard deviation of $\Theta_{\Lambda}$ will
converge to $0$. However, since $\Lambda _{T+1}$ viewed from time
$T$ is always random, the posterior standard deviation for
$\Lambda_{T+1}$ will be finite.}

\item \textnormal{Note that, conditionally given $\Theta_{\Lambda} =\theta_{\Lambda} $, $N_{t}$
has a negative binomial distribution with probability weights for
$n\geq 0$,
\begin{equation}
P\left. \left[ N_{t}=n\right\vert \theta_{\Lambda} \right] ={\binom{{\alpha +n-1}}{{n}}%
}\left( \frac{\alpha }{\alpha +\theta_{\Lambda} V}\right) ^{\alpha }\left( \frac{%
\theta_{\Lambda} V}{\alpha +\theta_{\Lambda} V}\right) ^{n}.
\end{equation}%
That is, at this stage we could directly work with a negative
binomial distribution. As we will see below, only in the marginal
case can we work with (4.4). In the multidimensional model we require $\Lambda _{t}$}$%
\mathnormal{.}$

\item \textnormal{$\Delta _{k}$ denotes the expert opinion of expert $k$
which predicts the true risk characteristics $\Theta_{\Lambda} $
of his company. We have
\begin{eqnarray}
E\left. \left[ \Delta _{k}\right\vert \Theta_{\Lambda} \right]
&=&E\left. \left[ \Lambda _{j}\right\vert \Theta_{\Lambda} \right]
=E\left. \left[ N_{j}/V\right\vert
\Theta_{\Lambda} \right] =\Theta_{\Lambda} ,  \notag \\
Var\left. \left( \Delta _{k}\right\vert \Theta_{\Lambda} \right) &=&\Theta_{\Lambda} ^{2}/\xi ,~~%
\text{ }CV\left. \left( \Delta _{k}\right\vert \Theta_{\Lambda}
\right) =\xi ^{-1/2}.
\end{eqnarray}%
That is, the relative uncertainty CV in the expert opinion does
not depend on the value of $\Theta_{\Lambda} $. That means that
$\xi $ can be given externally, e.g. by the regulator, who is able
to give a lower bound to the uncertainty. Moreover, we see that
the expert predicts the average frequency for his company.
Alternatively, $\xi$ can be estimated using method of moments as
presented in Lambrigger, Shevchenko and W\"{u}thrich (2007).}
\end{itemize}
\end{rems}

\noindent Denote
$\Lambda_{1:T}=\left(\Lambda_1,\ldots,\Lambda_T\right)$,
$N_{1:T}=\left(N_1,\ldots,N_T\right)$ and
$\Delta_{1:K}=\left(\Delta_1,\ldots,\Delta_K\right)$. Then the
joint posterior density of the random vector $(\Theta_{\Lambda}
,\Lambda _{1:T})$ given observations $N_{1}=n_{1},\ldots
,N_{T}=n_{T}$, $\Delta _{1}=\delta _{1},\ldots ,\Delta _{K}=\delta
_{K}$ is by Bayes' Theorem%
\begin{eqnarray}
\pi (\theta_{\Lambda} ,\lambda _{1:T}|n_{1:T},\delta_{1:K})
\propto \pi \left( n_{1:T}|\theta_{\Lambda} ,\lambda _{1:T}\right)
\pi \left( \lambda _{1:T}|\theta_{\Lambda} \right) \pi \left(
\delta _{1:K}|\theta_{\Lambda} \right) \pi \left( \theta_{\Lambda}
\right) .
\end{eqnarray}%
Here, the likelihood terms and the prior are made explicit,
\begin{eqnarray}
\pi \left( n_{1:T}|\theta_{\Lambda} ,\lambda_{1:T} \right) \pi
\left( \lambda_{1:T}|\theta_{\Lambda} \right) &=&\prod_{t=1}^{T}\frac{(V\lambda _{t})^{n_{t}}}{n_{t}!}%
\frac{(\alpha/\theta_{\Lambda})^{\alpha }}{\Gamma (\alpha
)}\lambda _{t}^{\alpha -1}\exp \left\{ -\lambda _{t}\left(V +
\alpha/\theta_{\Lambda}\right) \right\} ,
\\
\pi \left( \delta _{1:K}|\theta_{\Lambda} \right) &=&\prod_{k=1}^{K}%
\frac{(\xi/\theta_{\Lambda})^{\xi }}{\Gamma (\xi )}~\delta
_{k}^{\xi -1}~\exp
\left\{ -\delta _{k}\xi/\theta_{\Lambda}\right\} , \\
\pi \left( \theta_{\Lambda} \right) &=&\frac{b^{a}}{\Gamma
(a)}~\theta_{\Lambda} ^{a-1}~\exp \left\{ -\theta_{\Lambda}b
\right\} .
\end{eqnarray}%
Note that the intensities $\Lambda _{1},\ldots ,\Lambda _{T}$ are
non-observable. Therefore we take the integral over their
densities to obtain the posterior distribution of the random
variable $\Theta_{\Lambda} $ given $\left(N_{1:T},\Delta
_{1:K}\right)$

\begin{align}
\pi \left( \theta_{\Lambda} |n_{1:T},\delta _{1:K}\right) \propto & \prod_{t=1}^{T}{\binom{{\alpha +n_{t}-1}}{{n_{t}}}}\left( \frac{%
\alpha }{\alpha +\theta_{\Lambda} V}\right) ^{\alpha }\left(
\frac{\theta_{\Lambda} V}{\alpha+\theta_{\Lambda} V}\right) ^{n_{t}}  \nonumber \\
\ \ &\times \prod_{k=1}^{K}\frac{(\xi/\theta_{\Lambda})^{\xi }}{%
\Gamma (\xi )}~\delta _{k}^{\xi -1}~\exp \left\{ -\delta _{k}\xi/
\theta_{\Lambda}\right\} \frac{b^{a}}{\Gamma (a)}~\theta_{\Lambda}
^{a-1}~\exp \left\{ -\theta_{\Lambda}b\right\}  \nonumber \\
\ \ \propto & \left( \frac{1}{\alpha +\theta_{\Lambda} V}\right)
^{T\alpha +\sum_{t=1}^{T}n_{t}}~\theta_{\Lambda} ^{a-K\xi
+\sum_{t=1}^{T}n_{t}-1}~\exp \left\{ -\theta_{\Lambda}b
-\frac{\xi} {\theta_{\Lambda}}\sum_{k=1}^{K}\delta _{k}\right\} .
\end{align}

Given $\Theta_{\Lambda} $, the distribution of the number of losses
$N_t$ is negative binomial. Hence, one could start with a negative
binomial model for $N_t$. The reason for the introduction of the
random intensities $\Lambda _{t}$ is that we will utilize
them to model dependence between different risk cells, by introducing dependence between $%
\Lambda _{t}^{(1)},\ldots,\Lambda _{t}^{(J)}$.

Typically, a closed form expression for the marginal posterior function of $%
\Theta_{\Lambda} $, given $\left(N_{1:T},\Delta _{1:K}\right)$ can
not be obtained, except in this single risk cell setting. In
general, we will integrate out the latent variables $\Lambda
_{1},\ldots ,\Lambda _{T}$ numerically through a MCMC approach to
obtain an empirical distribution for the posterior of $\pi \left(
\theta_{\Lambda} |n_{1:T},\delta _{1:K}\right)$. This empirical
posterior distribution then allows for the simulation of $\Lambda _{T+1}$ and $N_{T+1}$%
, respectively, conditional on the observations $\left(N_{1:T},
\Delta_{1:K}\right)$.

\subsection{Modelling frequencies for multiple risk cells}
As in the previous section we will illustrate our methodology by
presenting the frequency model construction. In this section we will
extend the single risk cell frequency model to the general multiple
risk cell setting. This will involve formulation of the multivariate
posterior distribution.

\begin{model ass}[multiple risk cell frequency model]
\textrm{Consider J risk cells. Assume that every risk cell
}$j$\textrm{\ has a fixed, deterministic volume $V^{(j)}$.}

\begin{enumerate}
\item{ \textrm{The risk characteristic $\bm{\Theta}_{\Lambda} =(\Theta_{\Lambda} ^{(1)},\ldots
,\Theta_{\Lambda} ^{(J)})$ has a }$J$\textrm{-dimensional prior density }$\pi (%
\bm{\theta}_{\Lambda} )$. The copula parameters
$\bm{\theta}_{\rho}$ are modelled by a random vector
$\bm{\Theta}_{\rho}$ with the prior density
$\pi\left(\bm{\theta}_{\rho}\right)$; $\bm{\Theta}_{\Lambda}$ and
$\bm{\Theta}_{\rho}$ are independent.}

\item \textrm{Given} $\bm{\Theta}_{\Lambda} =\bm{\theta}_{\Lambda}$ and $\bm{\Theta}_{\rho}=\bm{\theta}_{\rho}$: $(\bm{\Lambda }_{1},\bm{N}%
_{1}),\ldots ,(\bm{\Lambda }_{T+1},\bm{N}_{T+1})$ \textrm{are
i.i.d.} \textrm{and the intensities} $\bm{\Lambda }_{t}$
$=(\Lambda
_{t}^{(1)},\ldots,\Lambda _{t}^{(J)})$ \textrm{have a } $J$ \textrm{%
-dimensional conditional density} with marginal distributions
$\Lambda _{t}^{\left( j\right) }\sim G\left( \cdot |\theta_{\Lambda}
^{\left( j\right) }\right) =\Gamma \left( \alpha ^{\left( j\right)
},\alpha ^{\left( j\right) }/\theta_{\Lambda} ^{\left( j\right)
}\right) $ and the copula $c(\cdot|\bm{\theta}_{\rho})$. Thus the
joint density of $\bm{\Lambda}_t$ is given by
\begin{equation}
\pi
(\bm{\lambda}_t|\bm{\theta}_{\Lambda},\bm{\theta}_{\rho})=c\left(
G(\lambda _{t}^{(1)}|\theta_{\Lambda} ^{(1)}),\ldots ,G(\lambda
_{t}^{(J)}|\theta_{\Lambda} ^{(J)})|\bm{\theta}_{\rho }\right)
\prod_{j=1}^{J}\pi (\lambda _{t}^{(j)}|\theta_{\Lambda} ^{(j)}),
\end{equation}%
where $\pi \left( \cdot |\theta_{\Lambda} ^{\left( j\right)
}\right) $ denotes the marginal density.

\item \textrm{Given $\bm{\Theta}_{\Lambda} =\bm{\theta}_{\Lambda} $ and $\bm{\Lambda }_{t}=%
\bm{\lambda }_{t}$, the number of claims are independent with
}\newline \textrm{$N_{t}^{(j)}\sim Poi(V^{(j)}\lambda
_{t}^{(j)}),j=1,\ldots ,J$. }

\item \textrm{There are expert opinions $%
\bm{\Delta }_{k}=(\Delta _{k}^{(1)},\ldots ,\Delta _{k}^{(J)})$,
$k=1,\ldots ,K$. Given $\bm{\Theta}_{\Lambda} =\bm{\theta}_{\Lambda}$: $\bm{\Delta }_{k}$ and $(%
\bm{\Lambda }_{t},\bm{N}_{t})$ are independent for all $k$ and
}$t$\textrm{; and $\Delta _{k}^{(j)}$ are all independent with $%
\Delta _{k}^{(j)}\sim \Gamma (\xi ^{(j)},\xi
^{(j)}/\theta_{\Lambda} ^{(j)})$. }
\end{enumerate}

\label{MultiVariateCaseModelAssumptions}
\end{model ass}
For convenience of notation, define:
\begin{itemize}
\item $\bm{\Lambda }_{1:T}=\left[ \left( \Lambda _{1}^{\left( 1\right)
},\ldots,\Lambda _{1}^{\left( J\right) }\right) ,\left( \Lambda
_{2}^{\left( 1\right) },\ldots,\Lambda _{2}^{\left( J\right)
}\right) ,\ldots,\left( \Lambda _{T}^{\left( 1\right)
},\ldots,\Lambda _{T}^{\left( J\right) }\right) \right] $ -
Frequency intensities for all risk profiles and years;

\item $\bm{N}_{1:T}=\left[ \left( N_{1}^{\left( 1\right) },\ldots,N_{1}^{\left(
J\right) }\right) ,\left( N_{2}^{\left( 1\right)
},\ldots,N_{2}^{\left( J\right) }\right) ,\ldots,\left(
N_{T}^{\left( 1\right) },\ldots,N_{T}^{\left( J\right) }\right)
\right] $ - Annual number of losses for all risk profiles and
years;

\item $\bm{\Delta }_{1:K}=\left[ \left( \Delta _{1}^{\left( 1\right)
},\ldots,\Delta _{1}^{\left( J\right) }\right) ,\left( \Delta
_{2}^{\left( 1\right) },\ldots,\Delta _{2}^{\left( J\right)
}\right) ,\ldots,\left( \Delta _{K}^{\left( 1\right)
},\ldots,\Delta _{K}^{\left( J\right) }\right) \right] $ - Expert
opinions on mean frequency intensities for all experts and risk
profiles.
\end{itemize}

\noindent \textbf{Prior Structure }$\pi \left(
\bm{\theta}_{\Lambda} \right)$ and $\pi \left( \bm{\theta}_{\rho}
\right).$ In the following examples, \textit{a priori,} the risk
characteristics $\Theta_{\Lambda} ^{(j)}$ are independent Gamma
distributed: $\Theta_{\Lambda} ^{(j)} \sim \Gamma
(a^{(j)},b^{(j)})$ with hyper-parameters $a^{(j)}>0$ and
$b^{(j)}>0$. This means that \textit{a priori} the risk
characteristics for the different risk classes are independent.
That is, if the company has a bad risk profile in risk class $j$
then the risk profile in risk class $i$ need not necessarily also
be bad. Dependence is then modelled through the dependence between
the intensities $\Lambda^{(1)}_{t},\ldots,\Lambda^{(J)}_{t}$. If
this is not appropriate then, of course, this can easily be
changed by assuming dependence within $\bm{\Theta}_{\Lambda} .$

In the simulation experiments below we consider cases when the
copula is parameterized by a scalar $\theta_{\rho}$. Additionally,
we are interested in obtaining inferences on $\theta_{\rho}$ implied
by the data only so we use an uninformative constant prior on the
ranges [-1,1], (0,30] and [1,30] in the case of Gaussian, Clayton
and Gumbel copulas respectively.

\noindent \textbf{Posterior density.} The marginal posterior density of
random vector $\left(\bm{\Theta}_{\Lambda},\Theta_{\rho}\right)$ given data of counts $\bm{N}_{1}=\bm{n}%
_{1},\ldots ,\bm{N}_{T}=\bm{n}_{T}$ and expert opinions $\bm{\Delta }_{1}=%
\bm{\delta }_{1},\ldots ,\bm{\Delta }_{K}=\bm{\delta }_{K}$ is

\begin{align}
\pi \left( \bm{\theta}_{\Lambda},
\theta_{\rho}|\bm{n}_{1:T},\bm{\delta}
_{1:K}\right) = & \prod_{t=1}^{T}\int \pi \left( \bm{\theta}_{\Lambda}, \theta_{\rho}, \bm{\lambda }_{t}|\bm{n}%
_{1:T},\bm{\delta }_{1:K}\right) d\bm{\lambda}_{t}  \nonumber \\
 \propto & \prod_{t=1}^{T}\left( \int ~\prod_{j=1}^{J}\exp \left\{
-V^{(j)}\lambda _{t}^{(j)}\right\} ~\frac{(V^{(j)}\lambda
_{t}^{(j)})^{n_{t}^{(j)}}}{n_{t}^{(j)}!}~\pi (\bm{\lambda }_{t}|\bm{\theta}_{\Lambda},\theta_{\rho})~d\bm{\lambda }_{t}\right)  \nonumber \\
&\times \prod_{k=1}^{K}\prod_{j=1}^{J}\left( \frac{(\xi
^{(j)}/\theta_{\Lambda} ^{(j)})^{\xi ^{(j)}}}{\Gamma (\xi
^{(j)})}~(\delta _{k}^{(j)})^{\xi ^{(j)}-1}~\exp \left\{ -\delta
_{k}^{(j)}\xi ^{(j)}/\theta_{\Lambda}
^{(j)}\right\} \right)  \nonumber \\
&\times \prod_{j=1}^{J}\frac{(b^{(j)})^{a^{(j)}}}{\Gamma
(a^{(j)})}~(\theta_{\Lambda} ^{(j)})^{a^{(j)}-1}~\exp \left\{
-b^{(j)}\theta_{\Lambda}
^{(j)}\right\}\pi\left(\theta_{\rho}\right) .
\label{PosteriorMultivariate}
\end{align}

\section{Simulation Methodology - Slice sampler}

Posterior (\ref{PosteriorMultivariate}) involves integration and
sampling from this distribution is difficult. Here we present a
specialized MCMC simulation methodology known as a Slice sampler to
sample from the desired target posterior distribution $\pi (
\bm{\theta}_{\Lambda}, \theta_{\rho}, \bm{\lambda }
_{1:T}|\bm{n}_{1:T}$, $\bm{\delta }_{1:K}) $. Marginally taken
samples of $\bm{\Theta}_{\Lambda}$ and $\Theta_{\rho}$ are samples
from $\pi \left( \bm{\theta}_{\Lambda},
\theta_{\rho}|\bm{n}_{1:T},\bm{\delta }_{1:K}\right) $ which can be
used to make inference for required quantities.

\noindent It will be convenient to define the exclusion operators,
$\bm{\Lambda}_{1:T}^{(-i,-j)}$, $\bm{\Lambda}_{1:T\backslash k}$ and
$\bm{\Theta}_{\Lambda}^{(-j)}$. For example:

\begin{itemize}
\item $\bm{\Lambda}_{1:T}^{\left(-2,-1\right) }=\left[ \left( \Lambda
_{1}^{\left( 1\right) },\ldots,\Lambda _{1}^{\left( J\right)
}\right) ,\left( \Lambda _{2}^{\left( 2\right) },\ldots,\Lambda
_{2}^{\left( J\right) }\right) ,\ldots,\left( \Lambda _{T}^{\left(
1\right) },\ldots,\Lambda _{T}^{\left( J\right) }\right) \right] $
- Frequency intensities for all risk profiles and years, excluding
risk profile 1 from year 2;

\item $\bm{\Lambda }_{1:T\backslash 2}=\left[ \left( \Lambda _{1}^{\left(
1\right) },\ldots,\Lambda _{1}^{\left( J\right) }\right) ,\left(
\Lambda _{3}^{\left( 1\right) },\ldots,\Lambda _{3}^{\left(
J\right) }\right) ,\ldots,\left( \Lambda _{T}^{\left( 1\right)
},\ldots,\Lambda _{T}^{\left( J\right) }\right) \right] $ -
Frequency intensities for all risk profiles and years, excluding
all profiles for year 2;

\item $\bm{\Theta}_{\Lambda}^{(-j)}=\left[ \Theta_{\Lambda}^{\left(
1\right) },\ldots,\Theta_{\Lambda}^{\left( j-1\right)
},\Theta_{\Lambda}^{\left( j+1\right)
},\ldots,\Theta_{\Lambda}^{\left( J\right) } \right].$

\end{itemize}

\noindent Sampling from $\pi \left( \bm{\theta}_{\Lambda},
\theta_{\rho},\bm{\lambda }_{1:T}|\bm{n}_{1:T},\bm{\delta
}_{1:K}\right) $ or $\pi \left( \bm{\theta}_{\Lambda},
\theta_{\rho}|\bm{n}_{1:T},\bm{\delta }_{1:K}\right) $ via closed
form inversion sampling or via rejection sampling is not typically
an option. There are many reasons for this. Firstly,
only for specific copula models will closed form tractable expression for the marginal $\pi \left( %
\bm{\theta}_{\Lambda},\theta_{\rho}|\bm{n}_{1:T},\bm{\delta
}_{1:K}\right) $ be attainable, certainly not for the models we
consider in this paper. Secondly, even for
the expression of the joint posterior $\pi \left( \bm{\theta}_{\Lambda}, \theta_{\rho}, \bm{\lambda }%
_{1:T}|\bm{n}_{1:T},\bm{\delta }_{1:K}\right) $ it is typically only
possible to sample from the conditional distributions sequentially
via numerical inversion sampling techniques which is highly
computational and inefficient in high dimensions. Additionally, we
would like a technique which is independent of the potentially
arbitrary choice in specifying a copula function for the dependence
between $\Lambda _{t}^{(1)},\ldots,\Lambda _{t}^{(J)}$. Hence, we
utilize an MCMC framework which we make general enough to work for
any choice of copula model, developed next.

\subsection{Bayesian Parameter Estimation}

We separate the analysis into two parts. Firstly, we condition on
knowledge of the copula parameters $\theta_{\rho}=\rho$, where
$\rho$ denotes the true copula parameters used to generate the
data. This allows us to demonstrate that if the copula parameters
$\theta_{\rho}$ are known, we can perform estimation of other
parameters accurately under joint inference. The second part
involves joint estimation of $\theta_{\rho}$ and
$\bm{\theta}_{\Lambda}$ to demonstrate the accuracy of the joint
inference procedure developed. Note that, the model for this
second part has not been formally introduced
but is a simple extension of Model Assumptions \ref%
{MultiVariateCaseModelAssumptions}.

\subsubsection{Conditional on a priori knowledge of copula parameter}

Here we assume the copula parameter has been estimated \textit{a
priori} and so estimation only involves model parameters. Such a
setting may arise for example if the copula parameter is already
estimated via a ML estimator. The proposed sampling procedure we
develop is a particular class of algorithms in the toolbox of MCMC
methods. It is an alternative to a Gibbs sampler known as a
univariate Slice sampler. We note that to implement a Gibbs sampler
or a univariate Slice sampler one needs to know the form of the full
conditional distributions. However, unlike the basic Gibbs sampler
the Slice sampler does not require sampling from these full
conditional
distributions.\ Derivations of the posterior full conditionals,%
\begin{eqnarray}
\pi (\theta_{\Lambda} ^{( j) }|\bm{\theta}_{\Lambda}^{( -j) }, \bm{\lambda }_{1:T},\bm{n}_{1:T},%
\bm{\delta }_{1:K},\theta_{\rho}) &\propto &\pi ( \bm{\lambda
}_{1:T}|\bm{\theta}_{\Lambda}^{( -j) },\theta_{\Lambda} ^{( j)
},\theta_{\rho}) \pi ( \bm{\delta }_{1:K}|\bm{\theta}_{\Lambda}^{(
-j) },\theta_{\Lambda} ^{(
j) })  \notag \\
&&\times \pi( \bm{\theta}_{\Lambda}^{( -j) }|\theta_{\Lambda} ^{(
j) }) \pi ( \theta_{\Lambda} ^{( j) }) ,
\label{postFullCondTheta} \\
\pi ( \lambda _{t}^{( j) }|\bm{\theta}_{\Lambda} ,\bm{\lambda}_{1:T}^{( -t,-j) },\bm{n}%
_{1:T},\bm{\delta }_{1:K},\theta_{\rho}) &\propto &\pi ( \bm{n}_{1:T}|\bm{\lambda }%
_{1:T}^{( -t,-j) },\lambda _{t}^{( j) }) \pi ( \bm{\lambda
}_{t}^{( -j) },\lambda _{t}^{(j) }|\bm{\theta}_{\Lambda}
,\theta_{\rho}) \label{postFullCondLambda}
\end{eqnarray}%
are presented in Appendix B.

\subsubsection{Joint Inference of marginal and copula parameters}
\label{extension1} To include the estimation of the copula
parameter $\theta_{\rho}$ jointly with the parameters
$\bm{\Theta}_{\Lambda}$ and latent intensities
$\bm{\Lambda}_{1:T}$ in our Bayesian framework, we assume that it
is constant in time and model it by a random variable
$\Theta_{\rho}$ with some prior density $\pi
(\theta_{\rho})$. The full conditional posterior of the copula parameter, denoted $\pi \left(\theta_{\rho} |%
\bm{\theta}_{\Lambda},\bm{\lambda}_{1:T},\bm{n}_{1:T},\bm{\delta}_{1:K}\right),$
is given by%
\begin{equation}
\pi \left(\theta_{\rho} |\bm{\theta}_{\Lambda},\bm{\lambda }_{1:T},\bm{n}_{1:T},\bm{\delta }%
_{1:K}\right) \propto \pi (\bm{\lambda
}_{1:T}|\bm{\theta}_{\Lambda} ,\theta_{\rho} )\pi \left(
\theta_{\rho} \right) , \label{postFullCondRho}
\end{equation}
For the full derivation of the scalar case see, Appendix B.

In the following section we provide intuition for our choice of
univariate Slice sampler as compared to alternative Markov chain
algorithms. In particular we describe the advantages that the Slice
sampler has compared to more standard Markov chain samplers, though
we also point out the additional complexity involved. We verify the
validity of the Slice sampling algorithm for those not familiar with
this specialized algorithm and we then describe some intricacies
associated with implementation of the algorithm. This is followed by
a discussion of some extensions we developed when analyzing the
OpRisk model. The technical details of the actual algorithm are
provided in Appendix C.

\subsection{Slice sampling}

The full conditionals given in equations (\ref{postFullCondTheta}), (\ref%
{postFullCondLambda}), (\ref{postFullCondRho}) do not take standard
explicit closed forms and typically the normalizing constants are
not known in closed form. Therefore this will exclude
straightforward inversion or basic rejection sampling being used to
sample from these distributions. Therefore one may adopt a
Metropolis Hastings (MH) within Gibbs sampler to obtain samples, see
for example Gilks, Richardson and Spiegelhalter (1996) and Robert
and Casella (2004) for detailed expositions of such approaches. To
utilize such algorithms it is important to select a suitable
proposal distribution. Quite often in high dimensional problems such
as ours, this requires tuning of the proposal for a given target
distribution. Hence, one incurs a significant additional
computational expense in tuning the proposal distribution parameters
off-line so that mixing of the resulting Markov chain is sufficient.
An alternative not discussed here would include an Adaptive
Metropolis Hastings within Gibbs sampling algorithm, see Atachade
and Rosenthal (2005) and Rosenthal (2009). Here we take a different
approach which utilizes the full conditional distributions, known as
a univariate Slice sampler, see Neal (2003). We demonstrate how
effective a univariate Slice sampler is for our model.

Slice sampling was developed with the intention of providing a
"black box" approach for sampling from a target distribution which
may not have a simple form. The Slice sampling methodology we
develop will be automatically tailored to the desired target
posterior. As such it does not require pretuning and in many cases
will be more efficient than a MH within Gibbs sampler. The reason
for this, pointed out by Neal (2003), is that a MH within Gibbs has
two potential problems. The first arises when a MH approach attempts
moves which are not well adapted to local properties of the density,
resulting in slow mixing of the Markov chain. Secondly, the small
moves arising from the slow mixing typically lead to traversal of a
region of posterior support in the form of a Random Walk. Therefore,
$L^{2}$ steps are required to traverse a distance that could be
traversed in only $L$ steps if moving consistently in the same
direction. A univariate Slice sampler can adaptively change the
scale of the moves proposed avoiding problems that can arise with
the MH sampler when the appropriate scale of proposed moves varies
over the support of the distribution.

A single iteration of the Slice sampling distribution for a toy
example is presented in Figure \ref{SliceSamplerFigure}. The
intuition behind Slice sampling arises from the fact that sampling
from a univariate distribution $p\left( \theta \right) $ can always
be achieved by sampling uniformly from the region under the
distribution $p\left( \theta \right) .$ Obtaining a Slice sample
follows two steps: sample a value $u_{l}\sim U\left[ 0,p\left( \theta_{l-1}\right) %
\right] $ and then sample a value uniformly from $A_{l},$ $\theta_{l}\sim U%
\left[ A_{l}\right] .$ This procedure is repeated and by
discarding the auxiliary variable sample $u_l$ one obtains
correlated samples $\theta_{l}^{\prime }s$ from $p\left(
\theta_{l-1}\right)$. Neal (2003), demonstrates that a Markov
chain $\left( U,\Theta \right) $ constructed in this way will have
stationary distribution defined by a uniform distribution under
$p\left( \theta \right) $ and the marginal of $\Theta $ has
desired stationary distribution $p\left( \theta \right) .$
Additionally, Mira and Tierney (2002) proved that the Slice sampler algorithm, assuming a bounded target distribution $%
p\left( \theta \right) $ with bounded support, is uniformly ergodic.

Similar to a deterministic scan Gibbs sampler, the simplest way to
extend the Slice sampler to a multivariate distribution is by
considering each full conditional distribution in turn. Note,
discussion relating to the benefits provided by Random Walk
behaviour suppression, as achieved by the Slice sampler, are
presented in the context of non-reversible Markov chains, see
Diaconis, Holmes and Neal (2000).

Additionally, we only need to know the target full conditional
posterior up to normalization, see Neal (2003) p. 710. This is
important in this example since solving the normalizing constant in
this model is not possible analytically. To make more precise the
intuitive description of the Slice sampler presented above, we
briefly detail the argument made by Neal on this point. Suppose we
wish to sample from a distribution for a random vector
$\mathbf{\Theta }\in \mathbb{R}^{n}$ whose density $p\left(
\bm{\theta }\right) $ is proportional to some function $f\left(
\bm{\theta }\right) $. This can be achieved by sampling uniformly
from the $\left( n+1\right) $-dimensional region that lies under the
plot of $f\left(\bm{\theta }\right) $. This is formalised by
introducing the auxiliary random variable $U$ and defining a joint
distribution over $\bm{\Theta}$ and $U$ which is uniform over the region $%
\left\{ \left( \bm{\Theta ,}U\right) :0<u<f\left( \bm{\theta }%
\right) \right\} $ below the surface defined by $f\left( \bm{\theta }%
\right) $, given by
\begin{equation*}
p\left( \bm{\theta ,}u\right) =\left\{
\begin{array}{cc}
{1/Z,} & \text{if }0<u<f\left( \bm{\theta }\right) , \\
{0,} & \text{ otherwise,}%
\end{array}%
\right.
\end{equation*}%
where $Z=\int f\left( \bm{\theta }\right) d\bm{\theta }$ . Then
the
target marginal density for $\mathbf{\Theta }$ is given by%
\begin{equation*}
p\left( \bm{\theta }\right) =\int_{0}^{f\left( \bm{\theta }\right) }%
\frac{1}{Z}du=\frac{f\left( \bm{\theta }\right) }{Z},
\end{equation*}%
as required. There are many possible procedures to obtain samples of
$\left( \mathbf{\Theta ,}U\right) $. The details of the implemented
algorithm undertaken in this paper are provided in Appendix C.
\bigskip

\noindent \textbf{Extensions}\newline We note that in the Bayesian
model we develop, in some cases strong correlation between the
parameters of the model will be present in the posterior, see Figure
\ref{fig2}. In more extreme cases, this can cause slow rates of
convergence of a univariate sampler to reach the ergodic regime,
translating into longer Markov Chain simulations. In such a
situation several approaches can be tried to overcome this problem.\
The first involves the use of a mixture transition kernel combining
local and global moves. For example, we suggest local moves via a
univariate Slice sampler and global moves via an Independent
Metropolis Hastings (IMH) sampler with adaptive learning of its
covariance structure, such an approach is known as a hybrid sampler,
see comparisons in Brewer, Aitken and Talbot (1996). Alternatively,
for the global move if determination of level sets in multiple
dimensions is not problematic, for the model under consideration,
then some of the multivariate Slice sampler approaches designed to
account for correlation between parameters can be incorporated, see
Neal (2003) for details. This is beyond the scope of this paper.

Another approach to break correlation between parameters in the posterior is
via transformation of the parameter space. If the transformation is
effective this will reduce correlation between parameters of the transformed
target posterior. Sampling can then proceed in the transformed space, and
then samples can be transformed back to the original space. It is not always
straightforward to find such transformations.

A third alternative is based on Simulated Tempering, introduced by
Marinari and Parisi (1992) and discussed extensively in Geyer and
Thompson (1995). In particular a special version of Simulated
Tempering, first introduced by Neal (1996) can be utilised in
which on considers a sequence of target distributions $\left\{ \pi
_{l}\right\} $ constructed such that they correspond to the
objective posterior in the following way,
\begin{equation*}
\pi _{l}=\left[ \pi \left( \bm{\theta}_{\Lambda},\bm{\lambda
}_{1:T},\theta_{\rho} \bm{|n}_{1:T},\bm{\delta }_{1:K}\right)
\right] ^{\gamma _{l}}
\end{equation*}%
with sequence $\left\{ \gamma _{l}\right\} .$ Then one uses the
Slice sampling algorithm presented and replaces $\pi$ with
$\pi_{l}$.

Running a Markov chain such that at each iteration $l$ we target
posterior $\pi _{l}$ and then only keeping samples from the Markov
chain corresponding to situations in which $\gamma _{l}=1$ can
result in a significant improvement in exploration around the
posterior support. This can overcome slow mixing arising from a
univariate sampling regime. The intuition for this is that for
values of $\gamma _{l}<<1$ the target posterior is almost uniform
over the space, resulting in large moves being possible around the
support of the posterior, then as $\gamma _{l}$ returns to a value
of $1$, several iterations later, it will be in potentially new
unexplored regions of posterior support.

As an extension we developed a Simulated Tempering Slice sampler
to obtain samples from the posterior
$p\left(\bm{\theta}_{\Lambda},\bm{\lambda}_{1:T},\theta_{\rho}
\bm{|n}_{1:T},\bm{\delta }_{1:K}\right)$. In our development we
utilize a sine function, $\gamma_l =
\text{min}\left(\text{sin}\left(\frac{2\pi}{1000}l\right)+1,1\right)$,
for $\gamma _{l}$ which has its amplitude truncated to ensure it ranges between $\left( 0,1%
\right] $. That is the function is truncated at $\gamma _{l}=1$
for extended iteration periods for our simulation index $l$ to
ensure the sampler spends significant time sampling from the
actual posterior distribution.

We note that application of the Tempering proved useful and
improved mixing of the Markov chain. However, for simulation
examples presented in the remainder of this paper it was
sufficient to use the basic univariate Slice sampler presented
previously, which is more computationally efficient than the
Tempered version.

Note, in the application of Tempering one must discard many
simulated states of the Markov chain, whenever $\gamma _{l}\neq
1$. There is however a computational way to avoid discarding these
samples, see Gramacy, Samworth and King (2007).

Finally, we note that there are several alternatives to a MH within
Gibbs sampler such as a basic Gibbs sampler combined with Adaptive
Rejection sampling (ARS), Gilks and Wild\textit{\ }(1992). Note ARS
requires distributions to be log concave. Alternatively an adaptive
version of this known as the Adaptive Metropolis Rejection sampler
could be used, see Gilks, Best and Tan (1995).

\section{Results}

In this section we demonstrate and compare the performance of our sampling
methodology on several different copula models. We intend to demonstrate the
appropriate behaviour of our Bayesian models as a function of the number of
annual years, in the presence of highly biased expert opinions. This will be
achieved through simulation studies using the sampling techniques detailed
above to perform inference on model parameters. The intention will be to
demonstrate the appropriate convergence and accuracy as a function of data
sample size. Hereafter, we study the case of dependence between intensities
of two risks and set risk cell volumes $V^{\left( 1\right) }=V^{\left(
2\right) }=1.$

\subsection{Estimation of model if copula parameter is known}

Here, we study the estimation of model parameters in two cases.
The first case involves two low frequency risks. In the second
case, one risk has low frequency while another risk has high
frequency. In these two cases we present results for the
univariate Slice sampler under scenarios involving: data generated
independently for each risk profile and data generated using a
Gaussian, Clayton and Gumbel copulas.

Only one expert opinion is assumed for each risk. We present the
parameter estimates as a function of data size for each of the
specified correlation levels. That is, we study the accuracy of the
parameter estimates as the number of observations increases.
Simulation results are obtained by creating independently 20 data
sets each of length 20 years, then for each data set simulations are
performed for subsets of the data going for 1, 2, 5, 10, 15 and 20
years. We then average the performance of posterior estimates over
these independent simulations. The Markov chains are run for 50,000
iterations with 10,000 iterations discarded as burnin. The
simulation time depends on the number of risk profiles, the number
of observations and expert opinions and the length of the Markov
chain\footnote{A typical run with 5 years of data and 1 expert in
the bivariate case for 50,000 simulations took approximately 50sec
and approximately 43min for the case of ten risk profiles when coded
in Fortran and run on 2.40GHz Intel Core2.}. In performing the
analysis we studied three cases and in each case we performed the
following steps,

\begin{enumerate}
\item Simulate a data set of appropriate number of years according to the
procedure specified in Appendix A.

\item Obtain correlated MCMC samples from the target posterior distribution
after discarding burnin samples, $\left\{ \bm{\theta }_{\Lambda, l},\bm{\lambda }%
_{1:T,l}\right\}, l=1001,\ldots,50000$.

\item Estimate desired posterior quantities such as posterior mean of
parameters of interest and posterior standard deviations.

\item Repeat stages 1 - 3 for 20 independent data realisations and then
average the results.
\end{enumerate}

\begin{itemize}
\item \textbf{Joint:} The results are obtained by MCMC samples taken from $\pi
\left(\bm{\theta}_{\Lambda}
,\bm{\lambda}_{1:T}|\bm{n}_{1:T},\bm{\delta
}_{1:K},\theta_{\rho}\right) $ with the correct copula model and
copula parameter used in the sampler. This is the procedure that
should be performed in a real application.

\item \textbf{Marginal:} Results are obtained by MCMC samples taken from
\begin{eqnarray*}
\pi \left( \bm{\theta}_{\Lambda} ,\bm{\lambda }_{1:T}|\bm{n}_{1:T},\bm{\delta }%
_{1:K},\theta_{\rho}\right) &=&\prod\limits_{j=1}^{J}\pi \left(
{\theta_{\Lambda} }^{\left( j\right) },{\lambda }_{1:T}^{\left(
j\right) }|\bm{n}_{1:T},\bm{\delta }_{1:K},\theta_{\rho}\right)
\\
&=&\prod\limits_{j=1}^{J}\pi \left( {\theta_{\Lambda} }^{\left( j\right) }|\bm{n}%
_{1:T},\bm{\delta }_{1:K},\theta_{\rho}\right)
\end{eqnarray*}%
which is the posterior in the case of independence. This is
equivalent to marginal estimation where single risk cell data is
analyzed separately, see Section 4.1.
\end{itemize}

\begin{itemize}
\item \textbf{Benchmark:} To verify the results we also consider the case
where we assume perfect knowledge of the realized random process
for random vector $\bm{\Lambda }_{1:T}$. We then perform inference
on $\bm{\Theta}_{\Lambda}$
without the additional uncertainty arising from estimating $\bm{\Lambda }%
_{1:T}.$ In this regard this represents a benchmark for which we may compare
the performance of our simulations. In particular, it is obtained by samples
taken from $\pi \left(\bm{\theta}_{\Lambda} |\bm{\lambda }_{1:T},\bm{n}_{1:T},%
\bm{\delta }_{1:K},\theta_{\rho}\right)$ conditional on the true
simulated realizations of random variables $\bm{\Lambda }_{1:T}$.
\end{itemize}

\noindent \textbf{Example 1: low frequency risk profiles.} We set
the true parameter values of $\Theta_{\Lambda}^{(1)}$ and
$\Theta_{\Lambda}^{(2)}$ to be $\theta _{true}^{\left( 1\right) }=5$
and $\theta _{true}^{\left( 2\right) }=5$ respectively. Also we
choose the expert's opinion on the true parameters to be an
underestimate in risk profile 1 with $\Delta_{1}^{\left( 1\right)
}=2$ and an overestimate for risk profile 2 with $\Delta
_{1}^{\left( 2\right) }=8.$ The model parameters were set to $\xi
^{\left( 1\right) }=\xi ^{\left( 2\right) }=2$, $\alpha ^{\left(
1\right) }=\alpha ^{\left( 2\right) }=2$, and prior distribution
parameters $a^{(1)}=a^{(2)}=2$, $b^{(1)}=b^{(2)}=2.5$. The results
for this simulation study, presented in Tables \ref{table1} and
\ref{table2}, show the appropriate convergence of the estimates of
parameters $\Theta_{\Lambda}^{(1)}$ and $\Theta_{\Lambda}^{(2)}$ as
a function of the data size, demonstrating how well this simulation
procedure works under these models. In addition we note that as
expected from credibility theory we observe that joint estimation is
better than the marginal, i.e. the posterior standard deviations for
$\Theta_{\Lambda}^{(1)}$ and $\Theta_{\Lambda}^{(2)}$ are less when
joint estimation is used. In addition the rate of convergence of the
posterior mean for $\Theta_{\Lambda}$ to the true value is faster
under the joint estimation. Note, the standard errors in the
posterior mean and standard deviation were calculated and found to
be strictly in the range of 1-5\% for the simulations presented.

In Figure \ref{fig2}, corresponding to Gaussian, Clayton and
Gumbel copula models respectively, we demonstrate the estimated
density $\pi \left( \bm{\theta}_{\Lambda} | \bm{\lambda
}_{1:T},\bm{n}_{1:T},\bm{\delta }_{1:K},\theta_{\rho}\right) $ if
we had perfect knowledge of the latent process parameters
$\bm{\Lambda }_{1:T}$. In this way we compare the exact posterior
with perfect knowledge of the correlation structure as captured by
the copula model which here we assume is known. Obtaining these
plots involved a particular realized data set of length 20 years.
For all copulas two values of $\theta_{\rho}$ were considered:
$\theta_{\rho}=0.9$ and $\theta_{\rho}=0.1$ for Gaussian copula;
$\theta_{\rho}=10$ and $\theta_{\rho}=1$ for Clayton copula; and
$\theta_{\rho}=3$ and $\theta_{\rho}=1.1$ for the Gumbel copula.
These plots of the joint marginal posterior distribution of
$\bm{\Theta}_{\Lambda}$ demonstrate clearly that the standard
practice in the industry of performing marginal estimation of risk
profiles will lead to incorrect results when estimating quantities
based on the distribution of $\bm{\Theta}_{\Lambda} $.

\bigskip

\noindent \textbf{Example 2: one low frequency and one high
frequency risk profile.} We set the true values of
$\Theta_{\Lambda}^{(1)}$ and $\Theta_{\Lambda}^{(2)}$ to be $\theta
_{true}^{\left( 1\right) }=5$ and $\theta _{true}^{\left( 2\right)
}=10$ respectively. Also we choose the expert's opinion on the true
parameters to be an under estimate in risk profile 1 with
$\Delta_{1}^{\left( 1\right) }=2$ and an over estimate for risk
profile 2 with $\Delta _{1}^{\left( 2\right) }=13.$ The model
parameters were set to $\xi ^{\left( 1\right) }=\xi ^{\left(
2\right) }=2$, $\alpha ^{\left(
1\right) }=2,\alpha ^{\left( 2\right) }=2$, $a^{(1)}=a^{(2)}=2$, $%
b^{(1)}=2.5,b^{(2)}=5$. The simulation results and comparisons are
developed in the same approach as Example 1 and again the standard
errors in the estimates were in the range 1-5\%. The results can
be found in Tables \ref {table3} and \ref{table4}.

\subsection{Joint estimation of marginal and copula parameters}
\label{extension} Here we estimate
$\Theta_{\Lambda}^{(1)},\Theta_{\Lambda}^{(2)}$ and
$\Theta_{\rho}$ jointly. For this example, the model settings from
Example 1 were used and one data set of length 20 years randomly
generated was utilized.\ The simulation was performed by taking
150,000 iterations of the sampler and discarding the first 20,000
as burnin. Results for these simulations are contained in Table
\ref{table5}.

These results demonstrate that our model and estimation
methodology is successfully able to estimate jointly the risk
profiles and the correlation parameter. This is seen to be the
case for all the models we consider in this paper. It is also
clear that with few observations, e.g. $T\leq5$, and a vague prior
for the copula parameter, it will be difficult to accurately
estimate the copula parameter. This is largely due to the fact
that the posterior distribution in this case is diffuse.
Additionally, with a small amount of data it appears that
accurately estimating the copula parameter is most difficult in
the Gumbel model. However, as the number of observations increases
the accuracy of the estimate improves in all models and the
estimates are reasonable in the case of $15$ or $20$ years of
data. Additionally, we could further improve the accuracy of this
prediction if we incorporated expert opinions into the prior
specification of the copula parameter, instead of using a vague
prior.

Overall, we have demonstrated that combination of all the relevant
sources of data can be achieved under our model. Then with this
study we show that our sampling methodology has the ability to
estimate jointly all the model parameters including the copula
parameter. This is a key step forward in model development and
estimation for OpRisk models. We further envisage that one can
extend this methodology to more sophisticated and flexible copula
based models with more than one parameter. This should be
relatively trivial since the methodology we developed applies
directly. However, the challenge in the case of a more
sophisticated copula model relates to finding a relevant choice of
prior distribution on the correlation structure.

~

\noindent\textbf{Full predictive distribution.} As a final comment
in this section we point out an important additional outcome of
obtaining samples from the joint posterior distribution of the model
parameters and the correlation. This relates to construction of the
full predictive annual loss distribution, accounting for parameter
uncertainty.

Typically practitioners will take point estimates of all
parameters and then condition on these point estimates to
empirically construct the predictive distribution and then
calculate risk measures to be reported such as VaR. Here we
comment that a more robust approach to prediction can now be
performed. Using our methodology, we can construct the full
predictive distribution after removing the parameter uncertainty
from the model considered, including the uncertainty arising from
the correlation parameter. To achieve this we would consider the
full predictive distribution:
\begin{equation}
\pi\left(Z_{T+1}|n_{1:T},\delta_{1:T}\right) = \int
\pi\left(Z_{T+1}|\bm{\theta}_{\Lambda},\theta_{\rho}\right)\pi\left(\bm{\theta}_{\Lambda},\theta_{\rho}
|n_{1:T},\delta_{1:T}\right)d\bm{\theta}_{\Lambda} d\theta_{\rho}.
\end{equation}
Here, we used the model assumptions that given
$\bm{\Theta}_{\Lambda}$ and $\Theta_{\rho}$ we have that $Z_{T+1}$
is independent from the observations $(N_{1:T},\Delta_{1:K})$. In
practice to obtain samples from this full predictive distribution
involves taking the steps demonstrated in Appendix A with a minor
modification. If one wanted to simulate $L$ annual losses from the
full predictive distribution, this would involve first running the
Slice sampler for $L$ iterations after burnin. Then for each
iteration $l$ one would use the state of the Markov chain
$\left(\bm{\theta}_{\Lambda,l},\theta_{\rho,l}\right)$ in the
simulation procedure detailed in Appendix A. We also note that it is
trivial under our methodology to extend this full predictive
distribution sampling to the case of frequency and severities.

\section{Discussion}

This paper introduced a dynamic OpRisk model which allows for significant
flexibility in correlation structures introduced between risk profiles. Next
a Bayesian framework was established to allow inference and estimation under
this model to be performed, whilst at the same time allowing incorporation
of alternative data sources into the inferential procedure. Then a novel
simulation procedure was developed for the Bayesian model presented, in the
case of dependence between frequency risk profiles. Simulations were
performed to demonstrate the accuracy of this procedure in multiple
bivariate examples. Comparisons were made between marginal estimation and a
benchmark estimation procedure. In all simulations, the estimation of the
model parameters was accurate and behaviour of the estimates of posterior
mean and standard deviation presented, smoothed over multiple data
realizations, was as expected. Initially the influence of the biased expert
opinion observation influenced the results and as the size of the data set
for actual annual loss counts grew, the estimations improved in accuracy.
Clearly, the joint estimation will outperform marginal estimation when
forming predictions of future counts and rates in year $T+1$, given
estimates based on data up to year $T$. Additionally, we demonstrated highly
accurate estimation of the copula parameter, jointly with the model
parameters.

Additionally, simulations were performed in the models $J=5$ and
$J=10$ for the Clayton copula model in which the copula parameter is
also unknown. Though the simulation time was increased as a factor
of the number of risk cells, the results and performance were as
presented for the bivariate models, making this approach suitable
for practical purposes.

Finally, the main objective of the paper is to preset the framework
for the multivariate problem and to demonstrate estimation in this
setting. Application of the framework to real data is the subject of
further research. In this paper, the estimation procedure is
presented for frequencies only but it can be extended in the same
manner as presented to severities.

\appendix

\section{Appendix: Simulation of annual losses}

\noindent In general, given marginal and copula parameters
$\left(\bm{\theta}_{\Lambda},\bm{\theta}_{\Psi},\bm{\theta}_{\rho}\right)$,
the simulation of the annual losses for year $t=T+1$, when risk
profiles are dependent, can be done as described below.

\begin{enumerate}
\item Simulate $2J$-variate $u_{1},\ldots,u_{J}, v_{1},\ldots,v_{J}$ from a $2J$
dimensional copula $C(\cdot|\bm{\theta}_{\rho}).$

\item Calculate ${\lambda }_{t}^{\left( j\right) }=G^{-1}\left( u_{j}|\theta_{\Lambda}^{(j)}\right)
$ and $\psi_{t}^{\left( j\right) }=H^{-1}\left(
v_{j}|\theta_{\Psi}^{(j)}\right), j=1,\ldots,J$

\item Sample $n_{t}^{(j)}$ from $P\left( \cdot |\lambda_{t}^{(j)}\right),
j=1,\ldots,J$.

\item Sample iid $x_{s}^{(j)}(t), s=1,\ldots,n_{t}^{(j)}, j=1,\ldots,J$ from $%
F\left(\cdot |{\psi }_{t}^{(j)}\right)$.

\item Calculate annual losses $z_{t}^{\left( j\right)
}=\sum\limits_{s=1}^{n_{t}^{\left( j\right) }}x_{s}^{(j)}\left(
t\right), j=1,\ldots,J$.

\item Repeat Steps 1-5, $K$ times to get $K$ random samples of the annual
losses $z_{t}^{\left( j\right) }$.
\end{enumerate}

\noindent Note, to simulate from the full predictive distribution of
annual losses, add simulation of
$\left(\bm{\theta}_{\Lambda},\bm{\theta}_{\Psi},\bm{\theta}_{\rho}\right)$
from the posterior distribution (e.g. using Slice sampler
methodology) as an extra step before Step 1. Simulation of the
random variates from a copula in Step 1 in the case of Gaussian,
Clayton and Gumbel copulas can be done as described below.

\noindent \textbf{Gaussian copula:}

\begin{enumerate}
\item Simulate $d$-variate $x_1,\ldots,x_d$ from $\Phi _{N}\left( \bm{0,}\Sigma
\right) $, where $\Phi _{N}\left( \bm{0,}\Sigma \right) $ is a Normal
distribution with zero means, unit variances and correlation matrix $\Sigma $%
.

\item Calculate $u_{1}=F_{N}(x_{1}),\ldots,u_d=F_{N}(x_{d})$. Obtained $%
(u_1,\ldots,u_d)$ is a $d$-variate from a Gaussian copula.
\end{enumerate}

\noindent \textbf{Archimedean copulas:} The Clayton and Gumbel copulas are
members of the Archimedean family of copulas. The $d$-dimensional
Archimedean copulas can be written as
\begin{equation}
C\left( u_{1},\ldots,u_{d}|\rho\right) =\phi ^{-1}\left( \phi
\left( u_{1}\right) +\cdots+\phi \left( u_{d}\right) \right)
\end{equation}
with $\phi $ a decreasing function known as the generator for the given
copula, see Frees and Valdez (1998). The generator and inverse generator for
the Clayton ($\phi _{C}$) and Gumbel ($\phi _{G}$) copulas are given by
\begin{eqnarray}
\phi _{C}\left( t\right) &=&\left( t^{-\rho }-1\right); \hspace{0.5cm} \phi
_{C}^{-1}\left( s\right) =\left( 1+s\right) ^{-\frac{1}{\rho }};  \notag \\
\phi _{G}\left( t\right) &=&\left( -\ln t\right) ^{\rho}; \hspace{0.5cm}
\phi_{G}^{-1}\left( s\right) =\exp \left( -s^{\frac{1}{^{\rho }}}\right),
\end{eqnarray}
where $\rho $ is a copula parameter. Simulation from such a copula
can be achieved following the algorithm provided in Melchiori
(2006):

\begin{enumerate}
\item Sample $d$ independent random variates $v_{1},\ldots,v_d$ from a uniform
distribution $U[0,1]$.

\item Simulate $y$ from $D(\cdot)$ such that Laplace transform of $D$
satisfies $\mathcal{L}\left( D\right) =\phi ^{-1}$ and $D\left(
0\right) =0.$

\item Find $s_{i}=-\left(\ln v_{i} \right) /y$ for $i=1,\ldots,d$

\item Calculate $u_{i}=\phi ^{-1}\left( s_{i}\right) $ for $i=1,\ldots,d.$
\end{enumerate}

\noindent The obtained $(u_{1},\ldots,u_{d})$ is a $d$-variate from
$d$-dimensional
Archimedean copula. What remains is to define the relevant distribution $%
D(\cdot)$ for the Clayton and Gumbel Copulas. For the Clayton copula, $%
D(\cdot)$ is a Gamma distribution with shape parameter given by
$\rho ^{-1}$
and unit scale. For the Gumbel copula, $D(\cdot)$ is from the $\alpha -$%
stable family $S_{\alpha }\left( \beta ,\gamma ,\delta \right) $ with the
following parameters shape $\alpha =\rho ^{-1}$, skewness $\beta =1$, scale $%
\gamma =( \cos(\frac{1}{2} \pi /\rho)) ^{\rho}$ and location
$\delta =0.$ In the Gumbel case, the density for $D$ has no
analytic form and the simulation from this distribution can be
achieved using the algorithm from Nolan (2007) to efficiently
generate the required samples from the univariate stable
distribution.

\section{Appendix: Full conditional posterior distributions}
Note, in Part 1 and Part 2 we are conditioning on the copula
parameter $\theta_{\rho}$, this notation is dropped for
simplicity. It is only explicitly introduced in Part 3.

\noindent\textbf{Part 1:} Using Bayes' theorem
\begin{align}
\pi & \left( \theta_{\Lambda} ^{\left( j\right) }|\bm{\theta}_{\Lambda}^{\left( -j\right) }, %
\bm{\lambda }_{1:T},\bm{n}_{1:T},\bm{\delta }_{1:K}\right)\propto \pi \left( %
\bm{\theta}_{\Lambda}^{\left( -j\right) },\bm{\lambda
}_{1:T},\bm{n}_{1:T},\bm{\delta }_{1:K}|\theta_{\Lambda} ^{\left(
j\right) }\right) \pi \left( \theta_{\Lambda} ^{\left(
j\right) }\right)  \notag \\
&=\pi \left( \bm{\lambda }_{1:T},\bm{n}_{1:T},\bm{\delta
}_{1:K}|\bm{\theta}_{\Lambda}
^{\left( -j\right) },\theta_{\Lambda} ^{\left( j\right) }\right) \pi \left( %
\bm{\theta}_{\Lambda}^{\left( -j\right) }|\theta_{\Lambda}
^{\left( j\right) }\right) \pi \left( \theta_{\Lambda} ^{\left(
j\right) }\right).
\end{align}
Using the model structure to exploit conditional independence properties we
note that
\begin{align}
\pi & \left( \bm{\lambda }_{1:T},\bm{n}_{1:T},\bm{\delta }_{1:K}|\bm{\theta}_{\Lambda} %
^{\left( -j\right) },\theta_{\Lambda} ^{\left( j\right) }\right)=\pi \left( \bm{n}%
_{1:T},\bm{\lambda }_{1:T}|\bm{\theta}_{\Lambda}^{\left( -j\right)
},\theta_{\Lambda} ^{\left( j\right) }\right) \pi \left(
\bm{\delta }_{1:K}|\bm{\theta}_{\Lambda}^{\left(
-j\right) },\theta_{\Lambda} ^{\left( j\right) }\right)  \notag \\
&=\pi \left( \bm{n}_{1:T}|\bm{\lambda }_{1:T}\right) \pi \left(
\bm{\lambda }_{1:T}|\bm{\theta}_{\Lambda}^{\left( -j\right)
},\theta_{\Lambda} ^{\left( j\right) }\right) \pi \left(
\bm{\delta }_{1:K}|\bm{\theta}_{\Lambda}^{\left( -j\right)
},\theta_{\Lambda} ^{\left( j\right) }\right)
\end{align}
which specifies the full conditional distributions for the
$j^{th}$ component $\Theta_{\Lambda} ^{\left( j\right) }$ as
\begin{align}
\pi & \left( \theta_{\Lambda} ^{\left( j\right) }|\bm{\theta}_{\Lambda}^{\left( -j\right) }, %
\bm{\lambda }_{1:T},\bm{n}_{1:T},\bm{\delta }_{1:K}\right) \propto \pi \left( \bm{n}_{1:T}|\bm{\lambda }_{1:T}\right)  \notag \\
&\hspace{1.0cm} \times \pi \left( \bm{\lambda
}_{1:T}|\bm{\theta}_{\Lambda}^{\left( -j\right) },\theta_{\Lambda}
^{\left( j\right) }\right) \pi \left( \bm{\delta
}_{1:K}|\bm{\theta}_{\Lambda}^{\left(
-j\right) },\theta_{\Lambda} ^{\left( j\right) }\right) \pi \left( \bm{\theta}_{\Lambda} %
^{\left( -j\right) }|\theta_{\Lambda} ^{\left( j\right) }\right)
\pi \left( \theta_{\Lambda}
^{\left( j\right) }\right)  \notag \\
&\propto  \pi \left( \bm{\lambda }_{1:T}|\bm{\theta}_{\Lambda}
^{\left( -j\right) },\theta_{\Lambda} ^{\left( j\right) }\right) \pi
\left( \bm{\delta }_{1:K}|\bm{\theta}_{\Lambda}
^{\left( -j\right) },\theta_{\Lambda} ^{\left( j\right) }\right) \pi \left( %
\bm{\theta}_{\Lambda}^{\left( -j\right) }|\theta_{\Lambda}
^{\left( j\right) }\right) \pi \left( \theta_{\Lambda} ^{\left(
j\right) }\right) .
\end{align}

\noindent\textbf{Part 2:} The next full conditional distribution we must
specify is given by
\begin{align}
\pi & \left( \lambda _{t}^{\left( j\right)
}|\bm{\theta}_{\Lambda},\bm{\lambda } _{1:T}^{\left( -t,-j\right)
},\bm{n}_{1:T},\bm{\delta }_{1:K}\right)\propto
\pi \left( \bm{\theta}_{\Lambda},\bm{\lambda }_{1:T}^{\left( -t,-j\right) },\bm{n}%
_{1:T},\bm{\delta }_{1:K}|\lambda _{t}^{\left( j\right) }\right) \pi \left(
\lambda _{t}^{\left( j\right) }\right)  \notag \\
& = \pi \left( \bm{\lambda }_{1:T}^{\left( -t,-j\right) },\bm{n}_{1:T},%
\bm{\delta }_{1:K}|\bm{\theta}_{\Lambda} ,\lambda _{t}^{\left(
j\right) }\right) \pi \left( \bm{\theta}_{\Lambda} |\lambda
_{t}^{\left( j\right) }\right) \pi \left( \lambda _{t}^{\left(
j\right) }\right) .
\end{align}%
We then use conditional independence properties of the model to get%
\begin{align}
\pi & \left( \bm{\lambda }_{1:T}^{\left( -t,-j\right) },\bm{n}_{1:T},%
\bm{\delta }_{1:K}|\bm{\theta}_{\Lambda} ,\lambda _{t}^{\left(
j\right) }\right)
\notag \\
& = \pi \left( \bm{n}_{1:T}|\bm{\lambda }_{1:T}^{\left( -t,-j\right)
},\lambda _{t}^{\left( j\right) }\right) \pi \left( \bm{\lambda }%
_{1:T}^{\left( -t,-j\right) }|\bm{\theta}_{\Lambda} ,\lambda
_{t}^{\left( j\right) }\right) \pi \left( \bm{\delta
}_{1:K}|\bm{\theta}_{\Lambda} \right)
\end{align}
giving the full conditional we are interested in, up to proportionality,
\begin{align}
\pi & \left( \lambda _{t}^{\left( j\right) }|\bm{\theta}_{\Lambda},\bm{\lambda }%
_{1:T}^{\left( -t,-j\right) },\bm{n}_{1:T},\bm{\delta
}_{1:K}\right) \propto \pi \left( \bm{n}_{1:T}|\bm{\lambda
}_{1:T}^{\left( -t,-j\right) },\lambda _{t}^{\left( j\right)
}\right) \notag
\\
& \times \pi \left( \bm{\delta }_{1:K}|\bm{\theta}_{\Lambda} \right) \pi \left( \bm{\lambda }_{1:T}^{\left( -t,-j\right) }|%
\bm{\theta}_{\Lambda} ,\lambda _{t}^{\left( j\right) }\right) \pi \left( \bm{\theta}_{\Lambda} %
|\lambda _{t}^{\left( j\right) }\right) \pi \left( \lambda _{t}^{\left(
j\right) }\right) .
\end{align}
We now demonstrate that this expression simplifies significantly. We can
show that the terms $\pi \left( \bm{\lambda }_{1:T}^{\left( -t,-j\right) }|%
\bm{\theta}_{\Lambda} ,\lambda _{t}^{\left( j\right) }\right) \pi \left( \bm{\theta}_{\Lambda} %
|\lambda _{t}^{\left( j\right) }\right) \pi \left( \lambda _{t}^{\left(
j\right) }\right) $ simplify as follows:
\begin{align}
\pi & \left( \bm{\lambda }_{1:T}^{\left( -t,-j\right)
}|\bm{\theta}_{\Lambda} ,\lambda _{t}^{\left( j\right) }\right)
\pi \left( \bm{\theta}_{\Lambda} |\lambda _{t}^{\left(
j\right) }\right) \pi \left( \lambda _{t}^{\left( j\right) }\right)=\frac{%
\pi \left( \bm{\theta}_{\Lambda} ,\bm{\lambda }_{1:T}^{\left(
-t,-j\right) },\lambda _{t}^{\left( j\right) }\right) }{\pi \left(
\bm{\theta}_{\Lambda} ,\lambda _{t}^{\left( j\right) }\right) }
\pi \left( \bm{\theta}_{\Lambda} |\lambda _{t}^{\left( j\right)
}\right) \pi \left( \lambda _{t}^{\left( j\right)
}\right)  \notag \\
&= \frac{\pi \left( \bm{\lambda }_{1:T}^{\left( -t,-j\right)
},\lambda
_{t}^{\left( j\right) }|\bm{\theta}_{\Lambda} \right) \pi \left( \bm{\theta}_{\Lambda} \right) }{%
\pi \left( \bm{\theta}_{\Lambda} |\lambda _{t}^{\left( j\right)
}\right) \pi \left( \lambda _{t}^{\left( j\right) }\right) } \pi
\left( \bm{\theta}_{\Lambda} |\lambda _{t}^{\left( j\right)
}\right) \pi \left( \lambda _{t}^{\left( j\right) }\right) \notag
\\
&=\pi \left( \bm{\lambda }_{1:T}^{\left( -t,-j\right) },\lambda
_{t}^{\left( j\right) }|\bm{\theta}_{\Lambda} \right) \pi \left(
\bm{\theta}_{\Lambda}  \right) .
\end{align}
Finally, we are left with the full conditional distribution
\begin{align}
\pi & \left( \lambda _{t}^{\left( j\right) }|\bm{\theta}_{\Lambda}
,\bm{\lambda } _{1:T}^{\left( -t,-j\right)
},\bm{n}_{1:T},\bm{\delta }_{1:K}\right) \propto \pi \left(
\bm{n}_{1:T}|\bm{\lambda }_{1:T}^{\left( -t,-j\right) },\lambda
_{t}^{\left( j\right) }\right) \notag \\
&\hspace{1cm} \times \pi \left( \bm{\delta }_{1:K}|%
\bm{\theta}_{\Lambda} \right) \pi \left( \bm{\lambda }_{1:T}^{\left( -t,-j\right) }|%
\bm{\theta}_{\Lambda} ,\lambda _{t}^{\left( j\right) }\right) \pi \left( \bm{\theta}_{\Lambda} %
|\lambda _{t}^{\left( j\right) }\right) \pi \left( \lambda _{t}^{\left(
j\right) }\right)  \notag \\
&\propto \pi \left( \bm{n}_{1:T}|\bm{\lambda }_{1:T}^{\left(
-t,-j\right)
},\lambda _{t}^{\left( j\right) }\right) \pi \left( \bm{\lambda }%
_{1:T}^{\left( -t,-j\right) },\lambda _{t}^{\left( j\right) }|\bm{\theta}_{\Lambda} %
\right)  \notag \\
& \propto \pi \left( \bm{n}_{1:T}|\bm{\lambda }_{1:T}^{\left(
-t,-j\right)
},\lambda _{t}^{\left( j\right) }\right) \pi \left( \bm{\lambda }%
_{t}^{\left( -j\right) },\lambda _{t}^{\left( j\right)
}|\bm{\theta}_{\Lambda} \right) .
\end{align}

\noindent \textbf{Part 3:} The full conditional distribution for the copula
parameter is given by
\begin{align}
\pi & \left( \theta_{\rho} |\bm{\theta}_{\Lambda},\bm{\lambda
}_{1:T},\bm{n}_{1:T},\bm{\delta }
_{1:K}\right)\propto \pi \left( \bm{\theta}_{\Lambda} ,\bm{\lambda }_{1:T},\bm{n}%
_{1:T}, \bm{\delta }_{1:K}|\theta_{\rho} \right) \pi \left( \theta_{\rho} \right)  \notag \\
&\propto \pi \left( \bm{n}_{1:T}|\bm{\lambda }_{1:T}\right) \pi \left( %
\bm{\delta }_{1:K}|\bm{\theta}_{\Lambda} \right) \pi (\bm{\lambda }_{1:T}|\bm{\theta}_{\Lambda} %
,\theta_{\rho} )\pi \left( \bm{\theta}_{\Lambda} \right) \pi \left( \theta_{\rho} \right) \notag \\
&\propto \pi (%
\bm{\lambda }_{1:T}|\bm{\theta}_{\Lambda} ,\theta_{\rho} )\pi
\left( \theta_{\rho} \right) .
\end{align}

\section{Appendix: Slice sampler algorithm.}
Here, we provide the explicit details involved into implementation
of a Slice sampler algorithm within a Gibbs sampler framework
discussed in Section 5. The iterations of the Slice sampler are
denoted by simulation index $l\in \mathbb{N}$.

\noindent
\hrulefill%

\textbf{Slice sampling:}

\begin{enumerate}
\item Initialize $l=0$ the parameter vector $\left[ \bm{\theta }_{\Lambda,0},%
\bm{\lambda }_{1:T,0},\theta_{\rho,0}\right] $ randomly or
deterministically.

\item Repeat while $l \leq L$

\begin{enumerate}
\item Set $\left[ \bm{\theta }_{\Lambda,l},\bm{\lambda }_{1:T,l},\theta_{\rho,l}\right] = %
\left[ \bm{\theta }_{\Lambda,l-1},\bm{\lambda
}_{1:T,l-1},\theta_{\rho,l-1}\right] $

\item Sample $j$ uniformly from set $\left\{ 1,2,\ldots ,J\right\} $

Sample new parameter value $\widetilde{\theta}_{\Lambda}^{\left(
j\right) }$ from the full
conditional posterior distribution $\pi \left( \theta_{\Lambda} ^{\left( j\right) }|%
\bm{\theta }_{\Lambda,l}^{\left( -j\right) }\bm{,\lambda }_{1:T,l},\bm{n}_{1:T},%
\bm{\delta }_{1:K},\theta_{\rho,l}\right) $.

Set $\theta _{\Lambda, l}^{\left( j\right)
}=\widetilde{\theta}_{\Lambda} ^{\left( j\right) }$.

\item Sample $j$ uniformly from set $\left\{1,2,\ldots,J\right\} $ and $t$
uniformly from set $\left\{1,\ldots,T\right\} $

Sample new parameter value $\widetilde{\lambda} _{t}^{\left(
j\right) }$ from the full conditional posterior distribution $\pi
\left( \lambda _{t}^{\left( j\right) }|\bm{\theta }_{\Lambda,
l}\bm{,\lambda }_{1:T,l}^{\left( -t,-j\right)
},\bm{n}_{1:T},\bm{\delta }_{1:K},\theta_{\rho,l}\right) $.

Set $\lambda _{t,l}^{\left( j\right) }=\widetilde{\lambda}
_{t}^{\left( j\right) }$.

\item Sample new parameter value $\widetilde{\theta}_{\rho}$ from the full conditional
posterior distribution

$\pi \left( \theta_{\rho} |\bm{\theta }_{\Lambda, l}\bm{,\lambda }_{1:T,l},\bm{n}_{1:T},%
\bm{\delta }_{1:K}\right) $.

Set $\theta_{\rho,l}=\widetilde{\theta}_{\rho}$.
\end{enumerate}

\item $l=l+1$ and return to 2.
\end{enumerate}

\noindent
\hrulefill%

\begin{figure}[tbp]
\centerline{\includegraphics[width=12 cm,height=6
cm]{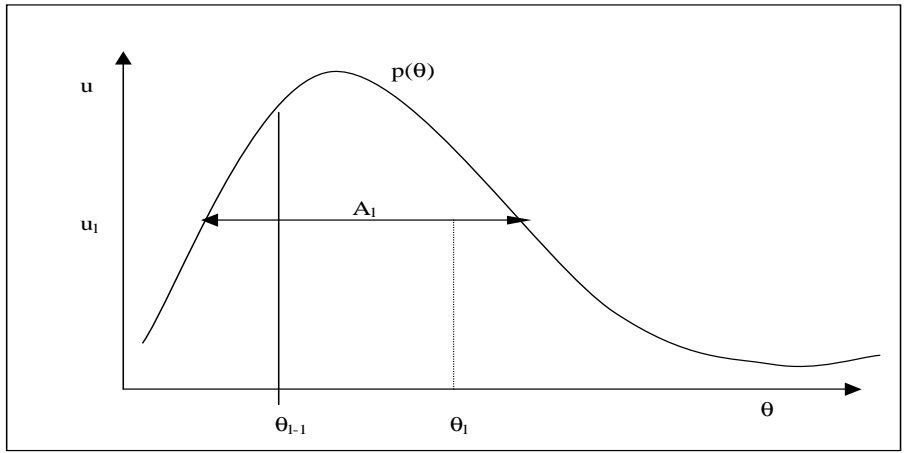}} \label{SliceSampler}
\caption{{\protect\footnotesize {Markov chain created for
$\Theta_{\Lambda} $ and auxiliary random variable $U$, \newline
$\left( u_{1},\protect\theta _{\Lambda, 1}\right) ,\ldots,\left(
u_{l-1},\protect\theta_{\Lambda, l-1}\right) ,\left(
u_{l},\protect\theta_{\Lambda,l}\right) ,\ldots$ has
stationary distribution with the desired marginal distribution $p\left( \protect%
\theta_{\Lambda} \right) .$}}} \label{SliceSamplerFigure}
\end{figure}

\begin{figure}[ptb]
\centerline{\includegraphics[width=6.0in,height=4.5in]{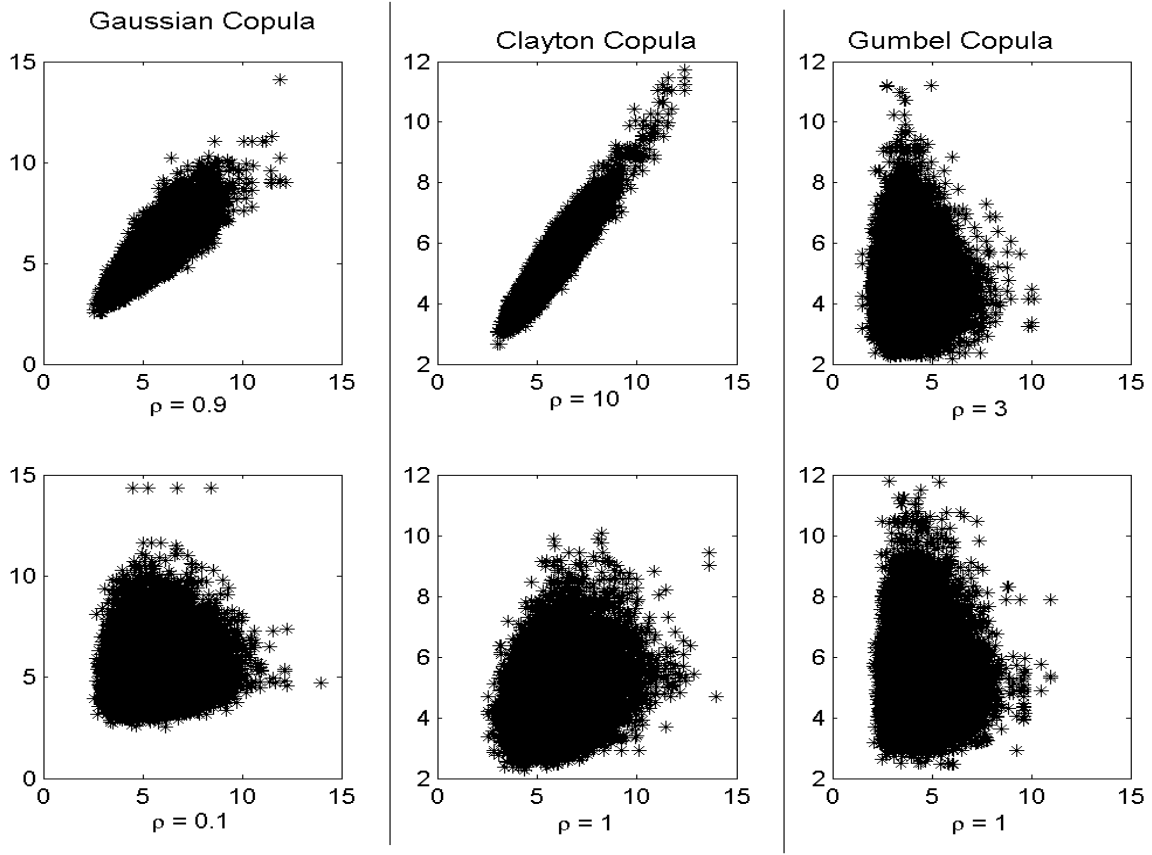}}
\caption{{\protect\footnotesize {Scatter plot of $\left(
\Theta_{\Lambda}
^{(1)},\Theta_{\Lambda}^{(2)}\right)$ from $\protect\pi\left( \bm{\theta}_{\Lambda} |\bm{n}%
_{1:20},\bm{\delta }_{1:1},\protect\lambda
_{1:20},\theta_{\rho}\right)$ with Gaussian, Clayton and Gumbel
copulas $C(\cdot|\protect\theta_{\rho}=\rho)$ between frequency
risk profiles. Top row: strong correlation. Bottom row: weak
correlation}}} \label{fig2}
\end{figure}

\noindent The sampling from the full conditional posteriors in stage
2 uses a univariate Slice sampler, see Figure
\ref{SliceSamplerFigure}. We present the case where we wish to
sample the next iteration of the Markov chain from $\pi \left(
\theta_{\Lambda} ^{\left( j\right)
}|\bm{\theta }_{\Lambda, l}^{\left( -j\right) },\bm{\lambda }_{1:T,l},\bm{n}_{1:T},%
\bm{\delta }_{1:K},\theta_{\rho}\right) .$

\noindent
\hrulefill%

\textbf{Obtaining a sample using a univariate Slice sampler:}

1.\qquad Sample $u$ from a uniform distribution%
\begin{equation*}
U\left[ 0,\pi \left( \theta _{\Lambda, l}^{\left( j\right)
}|\bm{\theta }_{\Lambda, l}^{\left(
-j\right) }\bm{,\lambda }_{1:T,l},\bm{n}_{1:T},\bm{\delta }_{1:K},\theta_{\rho}\right) %
\right] .
\end{equation*}

2.\qquad Sample $\widetilde{\theta}_{\Lambda} ^{\left( j\right) }$
uniformly from the
intervals (level set)%
\begin{equation*}
A=\left\{ \theta_{\Lambda} ^{\left( j\right) }:\pi \left( \theta_{\Lambda} ^{\left( j\right) }|%
\bm{\theta }_{\Lambda, l}^{\left( -j\right) }\bm{,\lambda }_{1:T,l},\bm{n}_{1:T},%
\bm{\delta }_{1:K},\theta_{\rho}\right) >u\right\} .
\end{equation*}

\noindent
\hrulefill%

\noindent There are many approaches that could be used in
determination of the level sets $A$ of our density
\begin{equation*}
\pi \left( \theta_{\Lambda} ^{\left( j\right) }|\bm{\theta }_{\Lambda, l}^{\left( -j\right) } ,%
\bm{\lambda }_{1:T,l},\bm{n}_{1:T},\bm{\delta
}_{1:K},\theta_{\rho}\right),
\end{equation*}%
see Neal (2003) [p.712, Section 4]. For simplicity in our proceeding
examples we assume that we can restrict our parameter space to the
finite ranges and we argue that this is reasonable since we can
consider the finite bounds for example set according to machine
precision for the smallest and largest number we can represent on
our computing platform. This is not strictly required, but
simplifies the coding of the algorithm. We then perform what Neal
(2003) terms a stepping out and a shrinkage procedure, the details
of which are contained in Neal (2003) [p.713, Figure 1]. The basic
idea is that given a sampled vertical level $u$ then the level sets
$A$ can be found by positioning an interval of width $w$ randomly
around $\theta_{\Lambda, l}^{\left(j\right)}$. This interval is
expanded in step sizes of width $w$ until both ends are outside the
slice. Then a new state is obtained by sampling uniformly from the
interval until a point in the slice $A$ is obtained. Points that
fail can be used to shrink the interval.


\newpage

\begin{table}[tbp]
\begin{center}
{\footnotesize
\begin{tabular}{|c|cccccc|}
\hline Year & 1 & 2 & 5 & 10 & 15 & 20 \\ \hline
\multicolumn{7}{|c|}{Independent} \\ \hline Marginal & 3.72 (2.04)
& 4.10 (1.98) & 4.08 (1.62) & 4.64 (1.42) & 5.13 (1.31) & 5.24
(1.27) \\ \hline \multicolumn{7}{|c|}{Gaussian copula with ($\rho
=0.9$)} \\ \hline Benchmark & 4.32 (1.88) & 4.50 (1.67) & 4.84
(1.46) & 5.17 (1.26) & 5.19
(1.12) & 5.21 (1.02) \\
Joint & 3.91 (2.01) & 4.41 (1.72) & 4.37 (1.56) & 4.76 (1.33) &
5.10 (1.21) & 4.95 (1.05) \\
Marginal & 3.72 (2.05) & 4.09 (1.97) & 4.06 (1.61) & 4.48 (1.37) & 5.07 (1.29) & 5.04 (1.13) \\
\hline \multicolumn{7}{|c|}{Clayton copula with ($\rho =10$)} \\
\hline Benchmark & 4.81 (1.82) & 5.17 (1.72) & 5.13 (1.42) & 4.96
(1.13) & 5.10 (0.98) & 5.00 (0.84) \\
Joint & 4.19 (2.03) & 4.92 (1.87) & 5.05 (1.56) & 4.87 (1.26) &
4.96 (1.08) & 4.90 (0.93) \\
Marginal & 3.91 (2.12) & 4.43 (2.10) & 4.54 (1.74) & 4.47 (1.36) & 4.75 (1.22) & 4.72 (1.08) \\
\hline \multicolumn{7}{|c|}{Gumbel copula with ($\rho =3$)} \\
\hline Benchmark & 4.32 (1.98) & 4.46 (1.70) & 4.86 (1.41) & 5.08
(1.16) & 5.16 (1.01) & 5.11 (0.88) \\
Joint & 4.33 (2.06) & 4.21 (1.80) & 4.54 (1.56) & 4.96 (1.23) &
5.01 (1.05) & 4.98 (0.93) \\
Marginal & 3.84 (2.08) & 3.76 (1.87) & 4.17 (1.62) & 4.63 (1.41) & 4.74 (1.22) & 4.72 (1.07) \\
\hline
\end{tabular}%
}
\end{center}
\caption{{\protect\footnotesize {Average estimates of posterior
mean and standard deviation of $\Theta_{\Lambda} ^{\left( 1\right)
}$ for 20 data sets. Data are generated using different copula models as specified. The true values are $\protect\theta %
_{true}^{\left( 1\right) }=\protect\theta _{true}^{\left( 2\right) }=5.$}}}
\label{table1}
\end{table}

\begin{table}[tbp]
\begin{center}
{\footnotesize
\begin{tabular}{|c|cccccc|}
\hline Year & 1 & 2 & 5 & 10 & 15 & 20 \\ \hline
\multicolumn{7}{|c|}{Independent} \\ \hline Marginal & 6.74 (2.74)
& 6.84 (2.59) & 6.46 (2.16) & 5.91 (1.67) & 5.74 (1.40) & 5.47
(1.31) \\ \hline \multicolumn{7}{|c|}{Gaussian ($\rho =0.9$)} \\
\hline Benchmark & 5.98 (2.29) & 5.84 (2.04) & 5.46 (1.60) & 5.47
(1.31) & 5.43
(1.14) & 5.41 (1.04) \\
Joint & 6.37 (2.55) & 6.01 (2.23) & 5.63 (1.80) & 5.40 (1.43) &
5.43 (1.25) & 5.36 (1.12) \\
Marginal & 6.59 (2.72) & 6.49 (2.54) & 6.01 (2.07) & 5.75 (1.64) & 5.62 (1.43) & 5.57 (1.26) \\
\hline \multicolumn{7}{|c|}{Clayton ($\rho =10$)}
\\ \hline Benchmark & 5.57 (1.91) & 5.41 (1.69) & 5.20 (1.40) &
4.90 (1.10) & 5.09
(0.96) & 5.07 (0.85) \\
Joint & 6.39 (2.48) & 5.92 (1.92) & 5.36 (1.64) & 5.06 (1.22) &
5.13 (1.17) & 5.00 (1.02) \\
Marginal & 6.69 (2.74) & 6.56 (2.55) & 5.92 (2.04) & 5.40 (1.56) & 5.37 (1.36) & 5.24 (1.17) \\
\hline \multicolumn{7}{|c|}{Gumbel ($\rho =3$)}
\\ \hline Benchmark & 5.83 (2.35) & 5.51 (2.02) & 5.38 (1.57) &
5.15 (1.18) & 5.20
(1.02) & 5.12 (0.89) \\
Joint & 6.05 (2.47) & 5.96 (2.17) & 5.47 (1.76) & 5.21 (1.27) &
5.12 (1.07) & 5.12 (0.94) \\
Marginal & 6.42 (2.67) & 6.26 (2.50) & 5.92 (2.04) & 5.67 (1.62) & 5.52 (1.37) & 5.36 (1.18) \\
\hline
\end{tabular}%
}
\end{center}
\caption{{\protect\footnotesize {Average estimates of posterior
mean and standard deviation of $\Theta_{\Lambda} ^{\left( 2\right)
}$ for 20 data sets. The data are generated using different copula models as specified. The true values are $\protect\theta %
_{true}^{\left( 1\right) }=\protect\theta _{true}^{\left( 2\right)
}=5$.}}} \label{table2}
\end{table}

\begin{table}[tbp]
\begin{center}
{\footnotesize
\begin{tabular}{|c|cccccc|}
\hline Year & 1 & 2 & 5 & 10 & 15 & 20 \\ \hline
\multicolumn{7}{|c|}{Independent} \\ \hline Marginal & 3.72 (2.04)
& 4.07 (1.97) & 4.05 (1.61) & 4.48 (1.37) & 4.94 (1.26) & 5.13
(1.13) \\ \hline \multicolumn{7}{|c|}{Gaussian ($\rho =0.9$)}
\\ \hline Benchmark & 4.07 (1.76) & 4.22 (1.53) & 4.61 (1.36) &
5.10 (1.20) & 5.10
(1.08) & 5.20 (0.99) \\
Joint & 3.86 (1.96) & 4.39 (1.86) & 4.46 (1.51) & 4.84 (1.33) &
5.10 (1.22) & 5.24 (1.10) \\
Marginal & 3.72 (2.04) & 4.08 (1.97) & 4.05 (1.61) & 4.48 (1.37) & 4.94 (1.26) & 5.13 (1.13) \\
\hline \multicolumn{7}{|c|}{Clayton ($\rho =10$)}
\\ \hline Benchmark & 4.45 (1.65) & 4.86 (1.55) & 4.89 (1.32) &
4.82 (1.07)
& 5.00 (0.95) & 4.92 (0.83) \\
Joint & 4.15 (2.01) & 4.84 (1.95) & 4.92 (1.59) & 4.69 (1.33) &
4.97 (1.20) & 4.96 (1.04) \\
Marginal & 3.98 (2.10) & 4.53 (2.09) & 4.54 (1.74) & 4.47 (1.36) & 4.74 (1.21) & 4.72 (1.08) \\
\hline \multicolumn{7}{|c|}{Gumbel ($\rho =3$)}
\\ \hline Benchmark & 4.14 (1.90) & 4.20 (1.58) & 4.65 (1.32) &
4.95 (1.11)
& 5.06 (0.97) & 5.04 (0.87) \\
Joint & 4.36 (2.16) & 4.17 (1.85) & 4.68 (1.57) & 5.10 (1.34) &
5.21 (1.21) & 5.24 (1.01) \\
Marginal & 3.84 (2.17) & 3.75 (1.87) & 4.17 (1.62) & 4.64 (1.41) & 4.75 (1.22) & 4.79 (1.09) \\
\hline
\end{tabular}%
}
\end{center}
\caption{{\protect\footnotesize {Average estimates of posterior
mean and standard deviation of $\Theta_{\Lambda} ^{\left( 1\right)
}$ for 20
data sets. Data are generated using different copula models as specified. The true values are $\protect\theta %
_{true}^{\left( 1\right) }=5$ and $\protect\theta _{true}^{\left(
2\right) }=10$.}}} \label{table3}
\end{table}

\begin{table}[tbp]
\begin{center}
{\footnotesize
\begin{tabular}{|c|cccccc|}
\hline Year & 1 & 2 & 5 & 10 & 15 & 20 \\ \hline
\multicolumn{7}{|c|}{Independent} \\ \hline Marginal & 10.89
(3.74) & 10.78 (3.60) & 10.18 (3.19) & 9.70 (2.67) & 9.64 (2.31) &
9.48 (2.14) \\ \hline \multicolumn{7}{|c|}{Gaussian ($\rho =0.9$)}
\\ \hline Benchmark & 10.44 (3.48) & 10.50 (3.24) & 10.25 (2.81) &
10.51 (2.39) &
10.78 (2.05) & 10.04 (1.88) \\
Joint & 10.68 (3.63) & 10.13 (3.35) & 9.57 (2.87) & 9.60 (2.36) &
9.31 (2.03) & 9.23 (1.82) \\
Marginal & 10.89 (3.74) & 10.78 (3.60) & 10.18 (3.19) & 9.70 (2.67) & 9.64 (2.31) & 9.48 (2.15) \\
\hline \multicolumn{7}{|c|}{Clayton ($\rho =10$)}
\\ \hline Benchmark & 10.04 (3.21) & 10.07 (2.99) & 9.88 (2.58) &
9.59
(2.08) & 9.97 (1.87) & 9.97 (1.66) \\
Joint & 10.58 (3.54) & 10.03 (3.19) & 9.29 (2.69) & 9.82 (2.22) &
9.93 (1.97) & 9.78 (1.73) \\
Marginal & 10.94 (3.75) & 10.92 (3.61) & 10.13 (3.17) & 9.32 (2.60) & 9.30 (2.25) & 9.45 (1.96) \\
\hline \multicolumn{7}{|c|}{Gumbel ($\rho =3$)}
\\ \hline Benchmark & 10.10 (3.55) & 9.88 (3.23) & 10.08 (2.74) &
9.98
(2.22) & 10.17 (1.95) & 10.06 (1.74) \\
Joint & 10.10 (3.61) & 10.18 (3.39) & 9.44 (2.87) & 9.98 (2.33) &
9.85 (1.99) & 9.63 (1.75) \\
Marginal & 10.61 (3.76) & 10.51 (3.59) & 10.11 (3.17) & 9.70 (2.66) & 9.45 (2.27) & 9.39 (1.98) \\
\hline
\end{tabular}%
}
\end{center}
\caption{{\protect\footnotesize {Average estimates of posterior
mean and standard deviation of $\Theta_{\Lambda} ^{\left( 2\right)
}$ for 20 data sets. Data are generated using different copula models as specified. The true values are $\protect\theta %
_{true}^{\left( 1\right) }=5$ and $\protect\theta _{true}^{\left(
2\right) }=10$.}}} \label{table4}
\end{table}

\begin{table}[tbp]
\begin{center}
{\footnotesize {\
\begin{tabular}{|c|cccccc|}
\hline
Year & 1 & 2 & 5 & 10 & 15 & 20 \\ \hline
\multicolumn{7}{|c|}{Posterior mean and standard deviation for $\Theta_{\Lambda}^{(1)}$} \\
\hline Independent & 2.81 (1.75) & 4.71 (2.21) & 3.05 (1.29) &
4.90 (1.48) & 4.38 (1.14) & 5.13 (1.15) \\
\hline Gaussian ($\rho =0.9$) & 2.83 (1.74) & 4.49 (2.02) & 3.31
(1.38) & 4.88 (1.29) & 4.36 (1.10) & 5.07 (1.09) \\
\hline Clayton ($\rho =10$) & 4.27 (2.04) & 3.80 (1.76) & 4.14
(1.46) & 5.62 (1.45) & 4.53 (1.04) & 4.90 (0.99) \\
\hline Gumbel ($\rho =3$) & 2.81 (1.40) & 2.59 (1.26) & 3.08 (1.17) & 4.28 (1.26) & 4.51 (1.12) & 4.91 (0.95) \\
\hline
\multicolumn{7}{|c|}{Posterior mean and standard deviation for $\Theta_{\Lambda}^{(1)}$} \\
\hline Independent & 10.24 (3.97) & 9.56 (3.56) & 9.48 (3.17) &
9.10 (2.58) & 9.33 (2.24) & 9.55 (1.98) \\
\hline Gaussian ($\rho =0.9$) & 10.23 (3.92) & 10.85 (3.52) & 8.72
(2.95) & 8.91 (2.12) &
8.58 (2.04) & 9.94 (1.85) \\
\hline Clayton ($\rho =10$) & 11.40 (3.58) & 10.76 (3.47) & 10.85
(3.10) & 11.39 (2.60) & 10.76 (2.27) & 10.17 (2.03)
\\ \hline
Gumbel ($\rho =3$) & 12.27 (3.63) & 11.26 (3.59) & 9.11 (2.93) & 9.54 (2.26) & 9.91 (1.72) & 9.88 (1.17) \\
\hline
\multicolumn{7}{|c|}{Posterior mean and standard deviation for $\Theta_{\rho} $} \\
\hline Independent & 0.20 (0.53) & 0.10 (0.44) & -0.10 (0.38) &
-0.02 (0.30) & -0.17 (0.28) & -0.12 (0.16) \\
\hline Gaussian ($\rho =0.9$) & 0.21 (0.54) & 0.47 (0.39) & 0.61
(0.30) & 0.66 (0.24) & 0.70 (0.19) & 0.74 (0.15) \\
\hline Clayton ($\rho =10$) & 5.37 (2.81) & 5.83 (2.66) & 6.20
(2.52) & 6.48 (2.25) & 6.70 (2.11) & 8.24 (1.88) \\
\hline Gumbel ($\rho =3$) & 16.41 (8.33) & 16.59 (8.19) & 16.39 (8.09) & 9.42 (8.76) & 5.12 (6.21) & 3.90 (4.80) \\
\hline
\end{tabular}
}}
\end{center}
\caption{{\protect\footnotesize {Posterior estimates for
$\Theta_{\Lambda} ^{\left( 1\right) },\Theta_{\Lambda} ^{\left(
2\right) }$ and copula parameter $\Theta_{\rho}$. In this case a
single data set is generated using different copula models as
specified. Posterior standard deviations are given in brackets
next to estimate. Joint estimation was used.}}} \label{table5}
\end{table}

\newpage

\end{document}